\documentclass[fleqn,onecolumn,usenatbib]{mnras}

\usepackage[T1]{fontenc}

\DeclareRobustCommand{\VAN}[3]{#2}
\let\VANthebibliography\thebibliography
\def\thebibliography{\DeclareRobustCommand{\VAN}[3]{##3}\VANthebibliography}

\usepackage{amsmath, amssymb}
\usepackage{bm}
\usepackage{graphicx}
\usepackage{subcaption}
\usepackage{enumitem}

\newcommand{\pD}[2]{\frac{\partial #2}{\partial #1}}

\newcommand{\D}[2]{\frac{{\rm d} #2}{{\rm d} #1}}

\newcommand{\bigD}[2]{\frac{{\rm D} #2}{{\rm D} #1}}
\newcommand\bb[1]{\mbox{\boldmath{$#1$}}}

\newcommand{\msb}[1]{\boldsymbol{\mathsf{#1}}}

\newcommand\mnz[1]{\left\langle {#1} \right\rangle_z}

\newcommand{\imag}{{\rm i}}
\newcommand{\rmd}{{\rm d}}
\newcommand{\rme}{{\rm e}}

\newcommand\bcdot{\,\bb{\cdot}\,}
\newcommand\btimes{\,\bb{\times}\,}
\newcommand\grad{\bb{\nabla}}
\newcommand{\ez}{\hat{\bb{z}}}
\newcommand{\ey}{\hat{\bb{y}}}
\newcommand{\ex}{\hat{\bb{x}}}
\newcommand{\eb}{\hat{\bb{b}}}
\newcommand{\vth}{v_{\rm th,i}}

\newcommand\bdbldot{\,\bb{:}\,}

\newcommand\mng[1]{\left\langle{#1}\right\rangle_\phi}

\def\apjs{Astrophys.~J.~Supp.~Ser.}
\def\apj{Astrophys.~J.}
\def\apjl{Astrophys.~J.~Lett.}
\def\aap{Astron.~Astrophys.}

\def\mnras{Mon.~Not.~Roy.~Astron.~Soc.}
\def\nat{Nature}

\def\prl{Phys.~Rev.~Lett.}
\def\prx{Phys.~Rev.~X.}
\def\jgr{J.~Geophys.~Res.}

\def\pof{Phys.~Fluids}

\def\jpp{J. Plasma Phys.}

\def\pasj{Pub.~Astron.~Soc.~Japan}

\usepackage{newtxtext,newtxmath}

\title[Collisionless spiral density waves]{Suppression of spiral density waves in collisionless accretion flows}

\author[Majeski, Kunz \& Dexter]{
Stephen Majeski,$^{1}$\thanks{E-mail: stephen.majeski@colorado.edu}
Matthew W. Kunz,$^{2,3}$
Jason Dexter$^{1,4}$
\\
$^{1}$JILA, University of Colorado and National Institute of Standards and Technology, 440 UCB, Boulder, CO 80309-0440, USA\\
$^{2}$Department of Astrophysical Sciences, Princeton University, Peyton Hall, Princeton, NJ 08544, USA\\
$^{3}$Princeton Plasma Physics Laboratory, PO Box 451, Princeton, NJ 08543, USA\\
$^{4}$Department of Astrophysical and Planetary Sciences, University of Colorado, 391 UCB, Boulder, CO 80309-0391, USA
}

\date{Accepted XXX. Received YYY; in original form ZZZ}

\pubyear{\the\year{}}

\begin{document}
\label{firstpage}
\pagerange{\pageref{firstpage}--\pageref{lastpage}}
\maketitle

\begin{abstract}
Spiral density waves are ubiquitous phenomena in differentially rotating flows. They are non-modal, non-axisymmetric perturbations that, due to shear in the background flow, possess time-evolving wavenumbers, frequencies, and amplitudes. In global simulations of optically thin, low-luminosity black-hole accretion flows, the excitation of such waves leads to spiral patterns in the plasma density that influence the morphology of the emission region. As such, the observed variabilities of such accretion flows are expected to be affected by the behavior of spiral density waves. Thus far, investigations of spiral density waves in accretion flows have exclusively used the assumption that the accretion-flow plasmas are in local thermodynamic equilibrium (LTE) at all times. In reality, hot and diffuse accretion-flow plasmas are expected to deviate far from LTE, and should be modeled accordingly. Here, we employ a combination of kinetic and fluid models to describe spiral density wave propagation, damping, and excitation in a collisionless shearing sheet, which more accurately represents the plasma conditions of low-luminosity accretion flows than do models that assume LTE. We find that micro-physical plasma instabilities are especially important in determining how spiral density waves behave, because the weak particle scattering that they induce leads to appreciable viscous damping. Observational implications of these results include reduced emission-region variability, which may improve agreement between Event Horizon Telescope (EHT) observations and simulation/theory, as well as a reduction in emission-region pattern speeds compared to simulations employing LTE, which bears on the interpretation of eventual EHT movies.
\end{abstract}

\begin{keywords}
accretion -- waves -- black hole physics -- plasmas -- magnetic fields
\end{keywords}

\section{Introduction}\label{sec:intro}

Non-axisymmetric spiral density waves are a common occurrence in differentially rotating flows. Well-documented examples range from spiral arms in galaxies, to perturbations driven by forming planets in protoplanetary disks, and even to waves in planetary rings~\citep[see, e.g.,][]{linshu64,perez16,goldreich78}. Such waves are also anticipated to arise in accretion flows around the supermassive black holes of galactic nuclei~\citep[e.g.,][]{jiang19,secunda19,conroy23,brown26}, although observations have yet to achieve the resolution needed to detect them directly. Their existence there has been predicted primarily using magnetohydrodynamic (MHD) simulation and theory: by global, general-relativistic calculations~\citep{eht22models} as well as local, non-relativistic ones~\citep[e.g., the shearing box;][]{hp09_1}. Both find the excitation of spiral density waves when the accretion flow is turbulent~\citep[e.g.,][]{hp09_2,fromang07,ross18}. Turbulence, in turn, has long been known to arise naturally from a host of instabilities that extract free energy from the shear of the rotating flow, the best known being the magnetorotational instability~\citep[MRI,][]{balbus91,bh91_2}. The MRI requires little more than a weakly magnetized plasma orbiting with an angular velocity that decreases outwards---conditions generic to nearly every class of accretion flow---and so both it and the turbulence it drives in its nonlinear phase are expected to be widespread. By association, spiral
density waves should be just as prevalent.

Recent work has brought the observational imprint of these waves into focus for black-hole accretion flows. The Event Horizon Telescope (EHT) collaboration compiled an immense library of global GRMHD simulations designed to replicate the observed near-horizon emission from the accretion flow onto the black hole at the center of our galaxy, Sgr~A$^\ast$~\citep{eht22models}. In virtually all of the simulated emission scenarios, spiral patterns were the dominant non-axisymmetric component of the synchrotron emission, indicating the presence of spiral density waves that had been excited farther out in the flow and propagated inwards. As discussed by \citet{conroy25}, this presents a unique opportunity: through these waves, the morphology and variability of emission in the region very near the black-hole event horizon may be directly connected to conditions farther out in the flow. Such a connection is especially valuable in light of the so-called ``variability crisis'': the fiducial models of the EHT library, as a whole, overestimated the variability of Sgr~A$^\ast$, and any individual models that did \textit{not} overestimate the variability were instead unable to satisfy other observational constraints. If spiral density waves are a significant contributor to that variability, the discrepancy implicates the modeling of the outer regions in which the waves are excited, rather than that of the near-horizon region where the emission is produced. This paper takes up that possibility by relaxing an assumption common to previous work---that the plasma is in local thermodynamic equilibrium (LTE)---an assumption with little justification in the hot and dilute plasmas of low-luminosity accretion flows, such as those around M87$^\ast$ and Sgr A$^\ast$. We posit that departures from LTE—driven by both the spiral density waves and the turbulence that excites them—modify the inward propagation of these waves, thereby linking the non-equilibrium thermodynamics of accretion flows to their observed properties and offering a potential route toward resolving the variability crisis of Sgr A$^*$.

\subsection{Spiral Wave Excitation in MRI Turbulence}\label{sec:hprev}

\begin{figure}
\centering
\includegraphics[width=0.72\linewidth]{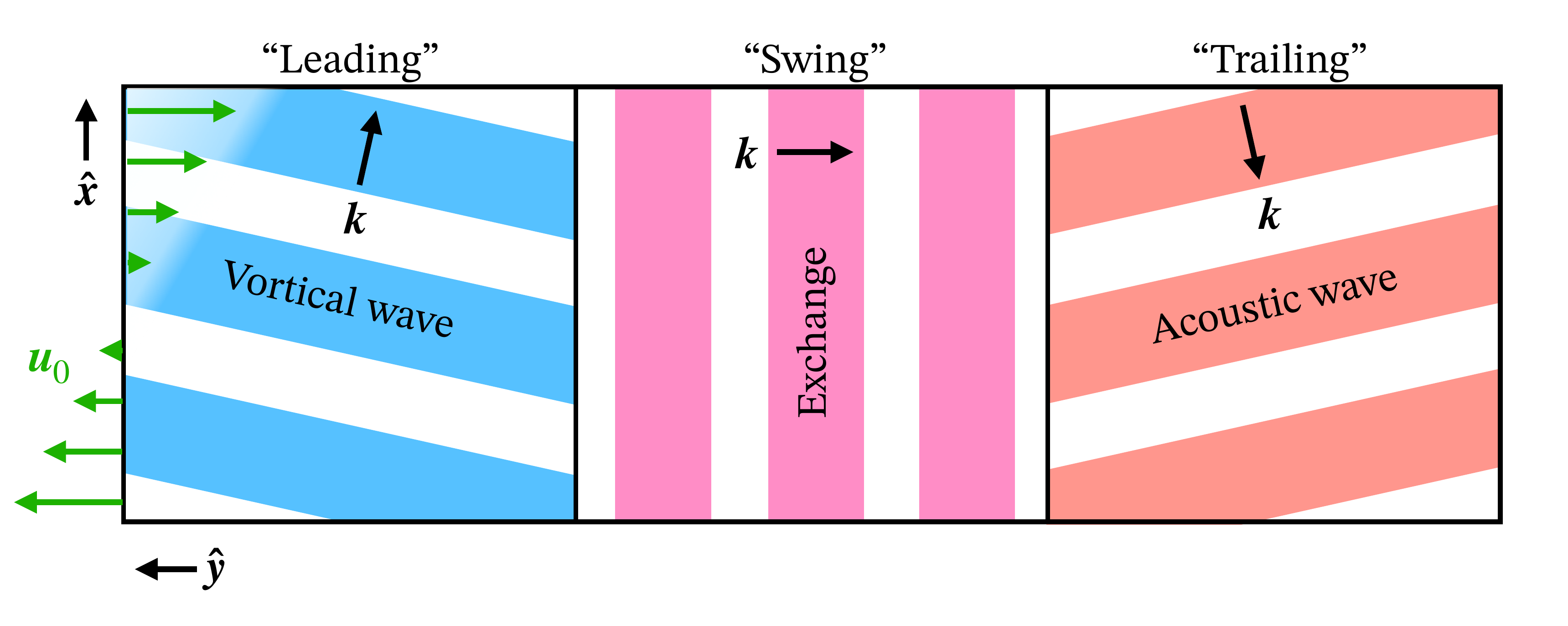}
\caption{Diagram showing the spontaneous excitation of density waves for a linearly sheared background flow $\bb{u}_0$. An initially leading vortical wave, which only weakly perturbs the density, is sheared by the flow. When it swings, all or some of its wave energy is converted into an acoustic wave, which strongly perturbs the density. The choice of axes mimics that of the shearing sheet, which we introduce in \S\ref{sec:eqs}.}
\label{fig:diag}
\end{figure}

Before considering how non-LTE effects may change spiral density waves, it is instructive to review first the mechanism that excites them. Although we focus solely on accretion flows in this work, the underlying physics of spontaneous density wave excitation actually does not require rotation at all. Instead, the essential ingredients are flow shear and fluctuations in the fluid vorticity. First described by \citet{chagelishvili97}, the basic process for density wave excitation in sheared flows is as follows (see figure \ref{fig:diag}). Initially, an aperiodic fluctuation in the fluid vorticity, known as a vortical wave, must be present. When strongly misaligned with the sheared flow, these waves perturb the density very weakly, thus, the initial conditions are nearly free of density perturbations. This vortical wave must also be ``leading'', which means that the ratio of its wavevector components perpendicular and parallel to the background flow $\bb{u}_0$ has the same sign as the gradient in $\bb{u}_0$ (e.g., $k_x/k_y < 0$ in figure \ref{fig:diag}). When subject to the sheared background flow $\bb{u}_0$, this leading wave experiences a rotation of its wavevector until it is completely aligned (or anti-aligned) with the flow direction, when it is said to ``swing'' (middle panel, figure \ref{fig:diag}). When a vortical wave swings, it becomes coupled to an acoustic wave, allowing energy to be exchanged between these two branches of the dispersion relation. This appears as a spontaneous excitation of density fluctuations, because energy is being pumped into a wave that strongly perturbs the density (the acoustic wave) from one that does not (the vortical wave). Following this conversion, the shear continues to rotate the resultant acoustic wave and the flow-perpendicular component of the wavevector changes sign from its initial condition to become ``trailing'' (right panel, figure \ref{fig:diag}). This trailing acoustic wave is a direct, non-rotating-flow analog of the spiral density waves that we study in this work.

The above conversion process was first generalized to include Keplerian rotation by \citet{bodo05}, however it was \citet{hp09_1} (hereafter H\&P) who demonstrated (both through analytic theory and simulation) that the excitation of spiral density waves should actually be commonplace in turbulent accretion flows. The authors found that MRI turbulence supplies the requisite fluctuations in vorticity, albeit in a slightly modified form specific to differentially rotating flows known as the pseudo-potential vorticity (or PPV). They demonstrated that fluctuations in the PPV grow slowly due to nonlinearities in MRI turbulence, and when those fluctuations swing due to the background flow shear, they are similarly converted into MHD fast waves, the magnetized-plasma equivalent of an acoustic wave. Just as in \citet{chagelishvili97}, this manifests as the sudden appearance of a propagating density fluctuation, which in an accretion flow takes on the form of a spiral wave. Using the ideal MHD equations, H\&P were able to describe the conversion process analytically, later verifying their results using MHD simulations in \citet{hp09_2}.

\subsection{Non-LTE Plasma Physics of Accretion Flows}\label{sec:nlte}

One shortcoming of prior studies of accretion-flow spiral density waves is that they assume that the plasma satisfies LTE at all times. Although the exact conditions of black-hole accretion flows are not observed directly, early analytic studies and the MHD-based simulations that followed have indicated that low-luminosity accretion flows are likely to be exceptionally hot (ion temperatures $T_{\rm i}\sim10^{9-12}$~K) and diffuse (ion densities $n_{\rm i}\sim10^{6-8}~\rm cm^{-3}$) in their equatorial plane~\citep[see, e.g.,][]{rees1982,narayan94,bh02,mocibrodzka09,pu16,chael18,white20}. In such plasmas, the relatively low densities and high temperatures make the collisional mean free path ($\lambda_{\rm mfp,i}$) much larger than the event horizon of the central supermassive black hole. In fact, given the above parameters, $\lambda_{\rm mfp,i}$ is \textit{at least} $50 \; \mu \rm pc \approx 10 \; au$, in excess of all characteristic scales of the accretion flow within ${\sim}100 r_g$ (the gravitational radius for a $4.3\times 10^6~{\rm M}_\odot$ black hole is $r_g \equiv 2GM_\bullet/c^2 \sim 0.1~{\rm au}$). Such a plasma can by no means be expected to remain in LTE, which is required for the MHD equations to be strictly valid. A non-LTE treatment of these regions may be needed to model spiral density waves in accretion flows accurately.

Spiral density waves are most likely to be affected by two non-LTE processes: resonant wave--particle interactions, and anisotropic-pressure stresses. Resonant wave--particle interactions arise because the phase-space distribution of particles (in position and velocity) is free to be rearranged by a wave. Individual particles that possess velocities slightly slower than a given wave's phase velocity are accelerated by its ponderomotive force, while those that move slightly faster are decelerated~\citep{vlasov38}. The tendency for stable distributions to be monotonic decreasing in energy ensures that more particles are moving slightly slower than the phase velocity, therefore these kinds of resonant interactions tend to extract net energy from the wave. This process is generally referred to as Landau damping~\citep{landau46}, whose most common manifestation results from the force of the wave's electric field. There are, however, alternative means by which a ponderomotive force can arise, such as through the magnetic mirror force of a compressive wave. The mirror force arises because collisionless, Larmor-gyrating particles have a conserved magnetic dipole moment, so any change in the magnetic-field strength effectively exerts a force on the particle's gyro-center in much the same way it would if the particle were replaced by a dipole magnet. In collisionless waves this force can lead to a form of damping often referred to as ``transit-time'' or ``Barnes'' damping~\citep{barnes66}. In general, any source of inertial or non-inertial field-aligned acceleration that yields a ponderomotive force can contribute to the resonant damping of a collisionless plasma wave, and the waves studied here will involve a combination of several ponderomotive forces. Importantly, however, all forms of Landau damping originate from a re-structuring of the distribution function, so none of them can be accurately captured by a fluid approach such as MHD. 

On the other hand, the consequences of pressure anisotropy can be captured by a fluid model, so long as a magnetic field is present to restrict momentum transport to occur predominantly along field lines. In such cases, the thermal pressure tensor is dominated by its components parallel to ($p_\|$) and perpendicular to ($p_\perp$) the local magnetic-field direction~\citep{cgl56}. This aspect alone leads to a restructuring of wave eigenmodes, due to the need to accommodate an additional evolution equation~\citep{hunana19,mks23}. These pressures are therefore also free to disagree with one another -- in fact, their very evolution is governed by conflicting invariants that guarantee pressure anisotropy ($\Delta p = p_\perp-p_\| \neq 0$) in the presence of density ($\rho$) or magnetic-field strength ($B$) fluctuations. The most well-known consequence of this pressure anisotropy is the viscous stress, which has long been understood as a source of dissipation in non-LTE plasmas. Therefore, like resonant interactions, pressure anisotropy and its associated viscous stress can lead to a damping of spiral density waves.

When the ratio of thermal pressure to magnetic pressure (known as $\beta_{(\perp,\|)} \equiv 8\pi p_{(\perp,\|)}/B^2$) is $\gtrsim 1$, $\Delta p$ takes on an additional role: a potent source of small-scale instability. Specifically, whenever $|\Delta p|$ becomes comparable to the energy density of the magnetic field, ion-Larmor-scale instabilities known as the mirror and firehose are triggered~\citep[e.g.,][]{parker58,barnes66}. Their thresholds for instability in a high-$\beta$ plasma are given by~\citep{sk93,bott24}
\begin{equation}\label{eq:thresh}
    \frac{\Delta p}{p_\|} \gtrsim \beta_\perp^{-1} \; {\rm (mirror)}, \quad \frac{\Delta p}{p_\|} \lesssim -1.4 \beta_\|^{-1} \; {\rm (firehose)}.
\end{equation}
While the underlying physics of these instabilities are distinct~\citep{rosin11,rincon15}, their consequences for the large-scale evolution of plasma thermodynamics are similar~\citep{kss14}. Both instabilities perturb the magnetic-field on sufficiently small scales that the Larmor motions of ions are disturbed. These perturbations induce pitch-angle scattering of particles with respect to the magnetic-field direction (at a rate we denote by $\nu_{\rm eff}$), which inevitably results in partial isotropization of the thermal pressures~\citep{schekochihin08,rosin11,kss14,riquelme15,rincon15,msk16}. In other words, a collisionless, high-$\beta$ plasma may be free of actual Coulomb collisions, but it may not be free of scattering. To avoid confusion then, in what follows we reserve the term ``collisions'' to describe Coulombic particle interactions exclusively, while ``scattering'' always refers to the isotropization of particle velocities caused by micro-instabilities. The rate at which micro-instability-induced scattering occurs is governed by the fact that it reduces the pressure anisotropy until $\Delta p$ is marginally unstable. \citet{kss14} found that a balance is ultimately established between the adiabatic driving and pitch-angle isotropization of $\Delta p$, which is achieved when the scattering rate satisfies
\begin{equation}\label{eq:scat}
\nu_{\rm eff} \sim \beta |3\eb\eb\bdbldot\grad \bb{u} - \grad \bcdot \bb{u}|.
\end{equation}
Here, $\bb u$ is the bulk flow velocity, $\eb$ is the magnetic-field unit vector, and we have used the fact that at high $\beta$, the thresholds \eqref{eq:thresh} become very stringent, so $\beta_\perp \approx \beta_\| \approx \beta$. In the presence of magnetic-field and density fluctuations that arise due to MRI turbulence, or even spiral density waves themselves, accretion flows may feature a significant amount of pressure anisotropy and therefore instability-mediated scattering~\citep{sharma06,kunz16,kempski19,dhruv25}. This scattering not only modifies the pressure anisotropy, but can also interrupt the wave--particle resonances that drive Landau/Barnes damping~\citep{kunz20,mks23}. Therefore, a careful investigation of spiral density waves in hot accretion flows must not only consider inherently large-scale effects such as resonant-wave particle interactions and viscous stresses, but also how micro-scale instabilities and their associated scattering modify those large-scale effects.

\subsection{Overview}

To reveal how non-LTE plasma physics affects the propagation, excitation, and damping of spiral density waves, we solve for the evolution of such waves in a local shearing-sheet model using semi-analytical techniques. In \S\ref{sec:eqs}, we introduce the governing equations used to incorporate the aforementioned non-LTE physics into our models, namely kinetic MHD~\citep[KMHD;][]{kulsrud83} and Braginskii-MHD~\citep{braginskii65}, both within the shearing-sheet approximation. Using these equations, in \S\ref{sec:lw} we solve for the linear evolution of spiral density waves in non-LTE accretion flows, beginning in \S\ref{sec:azimb} with the case of a laminar/azimuthal background magnetic field. Section~\ref{sec:azimb} is further divided into two scenarios depending on the presence or absence of scattering from micro-instabilities: in \S\ref{sec:nu0} we obtain spiral-wave solutions in both low- and high-$\beta$ plasmas assuming that no scattering of any kind is present, while in \S\ref{sec:nun0} we revisit the high-$\beta$ case including self-consistent scattering from micro-instabilities that are driven by the pressure anisotropy self-generated by the spiral waves. In \S\ref{sec:turbb} we study spiral waves in a high-$\beta$, turbulent plasma whose ``background'' magnetic field is isotropically tangled and the pressure anisotropy is dominated by the contributions from the turbulent shear and Keplerian shear.
We discuss the consequences of our findings in both \S\ref{sec:lw} and \S\ref{sec:disc}. Among them are the predictions that, compared to MHD, the excitation of non-LTE spiral waves in high-$\beta$, turbulent accretion flows is significantly suppressed by viscous dissipation (perhaps then contributing less to emission variability), and that the pattern speed of synchrotron structures in the emission region should be smaller than that predicted by MHD. Importantly, Landau/Barnes damping is \textit{not} found to be an efficient mechanism by which these spiral density waves can be dissipated in collisionless accretion flows. Instead, the dominant form of damping is viscous, a result of particle scattering induced by the mirror and firehose instabilities that are naturally driven by collisionless MRI turbulence. 
A consequence is that, in low-$\beta$ accretion flows where micro-instabilities are more difficult to excite and so the dominant form of density-wave damping is more likely to be Landau/Barnes damping, spiral density waves are not suppressed.
In \S\ref{sec:conc}, we summarize our investigation and consider next steps that should be taken.

\section{Model Equations in the Shearing Sheet}\label{sec:eqs}

Due to the added complexity of capturing relativistic effects near the central object, this work does not attempt a global model of spiral density waves in black-hole accretion flows. Instead, similarly to H\&P, we employ the non-relativistic shearing-sheet approximation to study the local excitation and damping of spiral waves \textit{prior} to their propagation into the emission region~\citep{glb65}. In the shearing sheet, the dynamical equations are transformed into a frame that rotates with the flow at a cylindrical-radial distance $r_0$ from the central object, which is much larger than the spatial extent of the domain being studied. This allows coordinate curvature to be neglected while incorporating the Coriolis force and a linear approximation to the difference between the centrifugal and gravitational forces. Rotation is taken to be about the $\ez$ axis at a steady rate $\Omega(r)$. The radial variation in $\Omega$ away from its value $\Omega_0$ at $r_0$ is approximated by a linear shear in the azimuthal flow, and, either by assumption or via spatial averaging, we do not consider variation in the $\ez$ direction---hence the name ``shearing sheet.'' The simplicity afforded by this local approach allows us to focus on the complexity added by the effects discussed in \S\ref{sec:nlte}, namely, anisotropic pressures, resonant wave--particle damping, and Larmor-scale instabilities. We now describe the two models that we use to capture the aforementioned effects: one appropriate to situations in which $\nu_{\rm eff}$ is not large enough to preclude resonant interactions (kinetic MHD), and one appropriate to situations when it is (Braginskii-MHD).

\subsection{Kinetic MHD}\label{sec:kmhd}

If the effective scattering frequency $\nu_{\rm eff}$ is smaller than or comparable to the characteristic dynamical frequency (\textit{viz.},~$\nu_{\rm eff}\lesssim v_{\rm th,i}/l$, with $l$ the characteristic dynamical length, $\vth= \sqrt{2T_{\rm i}/m_{\rm i}}$ the ion thermal speed, and $m_{\rm i}$ the ion mass), model equations must be used that account for collisionless effects such as resonant damping and non-LTE structure in velocity space. Kinetic magnetohydrodynamics~\citep[KMHD;][]{kulsrud83} provides such a set of equations, and applies when the dynamical timescales ($\omega^{-1}$) and lengthscales ($\ell$) are extremely large compared to those associated with the Larmor motions of individual charged particles, {\em viz.}~$\omega \ll \omega_{\rm L}$ and $l \gg r_{\rm L}$, where $\omega_{\rm L}$ and $r_{\rm L}$ are the Larmor frequency and radius. This well-magnetized limit is particularly suited for low-luminosity accretion flows. Consider Sgr~A$^\ast$ as an example~\citep{eht24polz}: the orbital frequency $\Omega$ (which approximates $\omega$) is ${\sim}10^{-3} \rm \; rad/s$ at $10 \,r_g$, while the Larmor frequency of ions at $10\, r_g$ is ${\sim}10^6 \; \rm rad/s$~\citep[assuming $B\approx 25 \rm \; G$; ][]{sen25}. This scale separation is satisfied even more handily for the electrons, however we do not model electrons through a kinetic approach in this work. In low-luminosity accretion flows, the electron temperature is thought to be much smaller than that of the ions~\citep{qg99,howes10,kawazura20}, so the electron pressure is negligible and their primary role is simply to screen the field-aligned component of the electric field (which in turn removes the ``classical'' channel of Landau damping, but does not affect others, such as transit-time damping). It is therefore unnecessary to solve for the electron distribution function here, so all references to a distribution function concern that of the ions alone.

In addition to the aforementioned length- and time-scale separation, another central feature of KMHD is that bulk plasma motions ($\bb{u}(t,\bb{r})$) are separated from individual particle velocities ($\bb{v}$) so that the resultant distribution function is solved for in the (non-inertial) frame of $\bb{u}$. This is accomplished by exchanging $\bb{v}$ in favor of the peculiar velocities $\bb{w} \equiv \bb{v} - \bb{u}$ in the distribution function $f(t,\bb{r},\bb{w})$. Importantly, $\bb{u}$ includes not only turbulence and waves, but also the background shear flow inherent to the shearing sheet approximation. Following this transformation, $f$ is then averaged over Larmor orbit gyro-phases to reduce the velocity-space dependence from $(w_x,w_y,w_z)$ to $(w_\perp,w_\|)$, where $\parallel$ and $\perp$ denote the directions along and across the local magnetic field. The evolution of the resultant gyrotropic distribution $f(t,\bb{r},w_\perp,w_\|)$ is given by the drift-kinetic equation
\begin{equation}\label{eq:kkin}
    \bigD{t}{f} + \bigD{t}{\ln B}\frac{w_\perp}{2}\pD{w_\perp}{f} - \biggl[ \eb\bcdot \biggl(\bigD{t}{\bb{u}} + 2\Omega_0\ez\btimes\bb{u} + 4A_0\Omega_0 x\ex\biggr)+\frac{w_\perp^2}{2}\eb\bcdot\grad\ln B \biggr]\pD{w_\|}{f} = \left( \frac{\partial f}{\partial t}\right)_{\!\rm scatt,eff} ,
\end{equation}
where ${\rm D}/{\rm D}t \equiv \partial / \partial t + \bb{u}\bcdot\grad +w_\|\eb\bcdot\grad$. Here, $\ex$ represents the local radial direction, $\ey$ the local azimuthal direction, $\ez$ the vertical direction, and $A_0$ (Oort's first constant) the rate of shear in the differential rotation local to $r_0$; $A_0=-(3/4)\Omega_0$ for a Keplerian flow, which we adopt uniformly throughout this work. The second term on the left-hand side of \eqref{eq:kkin} accounts for the change in the perpendicular velocity of the particles in the presence of a time-dependent magnetic-field strength due to conservation of particle magnetic moments. The third term (the one in square brackets) describes field-parallel forces on the particle guiding centers, including those that arise from boosting to the non-inertial reference frame of the bulk flow (a derivation of those terms is provided in \S\ref{app:kmhd}). Because $\bb{u}$ contains the background shear flow, which is given by $\bb{u}_0 = 2A_0x\ey$ in the shearing sheet, the terms $2\Omega_0\ez\btimes\bb{u}$ and $4 A_0 \Omega_0 x \ex$ cancel in equilibrium, so the only direct contributions of the shearing sheet to the drift-kinetic equation are via the perturbed Coriolis force and the presence of $\bb{u}_0$ in the convective derivative. We have also dropped the force due to the parallel electric field, which, as mentioned earlier, is a consequence of the electron temperature being negligible compared to that of the ions. The unspecified term on the right-hand side represents the effects of particle scattering caused by micro-instabilities on the evolution of $f$. 

The evolution of the bulk flow in KMHD is given by the MHD momentum equation with a pressure tensor that is anisotropic with respect to the local magnetic-field direction:
\begin{equation}\label{eq:kmom}
    \rho \D{t}{\bb{u}}=-\grad\left(p_{\perp}+\frac{B^2}{8\pi}\right)+\grad \bcdot\left[\eb\eb\biggl(p_\perp - p_\parallel + \frac{B^2}{4\pi}\biggr)\right]-2\rho\Omega_0\ez\btimes \bb{u}-4\rho A_0\Omega_0 x\ex,
\end{equation}
where $\rmd/\rmd t \equiv \partial/\partial t + \bb{u}\bcdot\grad$ is the convective derivative. Given our assumption of cold electrons, the $p_{(\perp,\|)}$ refer only to the anisotropic components of the ion pressure tensor. The magnetic-field induction equation is unmodified from ideal MHD:
\begin{equation}
    \D{t}{\bb{B}}= \bb{B} \bcdot \grad \bb{u} - \bb{B} \grad \bcdot \bb{u}.
\end{equation}
To close the KMHD system, the moments of the distribution $f$ are calculated using the peculiar velocities $(w_\perp,w_\|)$:
\begin{equation}\label{eq:moments}
    \rho = m_{\rm i}\int \rmd^3w\, f \,, \quad p_\perp = \frac{m_{\rm i}}{2}\int \rmd^3w \,w_\perp^2f \,, \quad p_\| = m_{\rm i}\int \rmd^3w \,w_\|^2f \,,
\end{equation}
where $\rmd^3w = 2\pi w_\perp \rmd w_\perp \rmd w_\|$. Because $f$ is specified in the frame of the local bulk-flow velocity, it has no velocity moments. 

The scattering term that we employ in this work for the right-hand side of \eqref{eq:kkin} is a version of the Krook operator that drives the perturbed distribution function toward a local Maxwellian having the same isotropic temperature and density as $f$~\citep{bgk56}:
\begin{equation}\label{eq:coll}
    \left( \frac{\partial f}{\partial t}\right)_{\!\rm scatt,eff} = - \nu_{\rm eff}\left[ f(t,\bb{r},w_\perp,w_\|) - F_{\rm M}(t,\bb{r},w) \right], \quad{\rm where}\quad F_{\rm M}(t,\bb{r},w) \equiv \frac{\rho(t,\bb{r})\sqrt{m_{\rm i}}}{(2\pi T_{\rm i}(t,\bb{r}))^{3/2}}\exp\left( -\frac{m_{\rm i}w^2}{2T_{\rm i}(t,\bb{r})}\right),
\end{equation}
$w^2\equiv w^2_\perp + w^2_\|$, and the temperature $T_{\rm i} = (m_{\rm i}/\rho) \,p = (m_{\rm i}/\rho) (2 p_\perp + p_\| )/3$. In the collisionless accretion flows studied here, the effective scattering rate, $\nu_{\rm eff}$, refers exclusively to scattering induced by micro-instabilities. One caveat of using \eqref{eq:coll} to mimic pitch-angle scattering from micro-instabilities, however, is that pitch-angle scattering only \textit{isotropizes} the distribution with respect to the magnetic field; it does not necessarily drive $f$ toward a Maxwellian. Nonetheless, when linearized, \eqref{eq:coll} affects all fluid moments of $f$ below the heat flux in a manner that exactly mimics pure pitch-angle scattering, so the consequences of this distinction are likely to be minimal in the context of the spiral waves studied here.\footnote{Alternatively, a more accurate choice, at least for the firehose instability, would be the effective firehose collision operator of \citet{bott25} (see their equation 6.1). Their operator is much more complex, however, and, compared to our simple Krook prescription, only changes the viscous stress by an order-unity amount.}

\subsection{The Braginskii-MHD Closure}\label{sec:brag}

Under certain circumstances, such as $\beta \gg 1$, the rate of scattering in \eqref{eq:scat} may be so large that the condition $l \gg v_{\rm th,i}/\nu_{\rm eff}$ is satisfied. In this case, Landau/Barnes damping may be ignored because resonant interactions are interrupted before they can damp the wave~\citep{zk63}, and the strong scattering keeps the pressure anisotropy much smaller than the isotropic pressure ($p$). If, additionally, the flow satisfies $1 \gtrsim u/\vth\gg \vth/(l\,\nu_{\rm eff})$, then $\Delta p$ can be approximated by the model of \citet{braginskii65}:
\begin{equation}\label{eq:brag}
\Delta p \approx \frac{3 p}{\nu_{\rm eff}} \left(\eb \eb- \frac{1}{3}\msb{I}\right)\bdbldot\grad \bb{u},
\end{equation}
where $\msb{I}$ is the unit dyadic. The corresponding isotropic component of the pressure is evolved according to
\begin{equation}\label{eq:isop}
\D{t}{p} = \frac{5}{3}\frac{p}{\rho}\D{t}{\rho} + \frac{2\nu_{\rm eff}}{9} \frac{(\Delta p)^2}{p} + \grad \bcdot \left[\frac{2\sigma }{3}\frac{p}{\nu_{\rm eff}}\,\eb\eb\bcdot \grad\left(\frac{p}{\rho}\right) \right],
\end{equation}
where the factor $\sigma$ depends on the form of the collision operator present in the kinetic equation. For the Krook operator that we employ \eqref{eq:coll}, $\sigma=5/2$; a more standard value is $\sigma=3.91$, which corresponds to the linearized form of the Landau collision operator~\citep{spitzer62,braginskii65}. From \eqref{eq:isop}, we see that the combination of pressure anisotropy and scattering leads to irreversible heating, and that conductive heat transport occurs only along the magnetic field. By eliminating the time derivative of the pressure anisotropy in favor of a steady-state balance between its adiabatic production and its relaxation due to scattering, the Braginskii approximation represents a considerable simplification from KMHD. The closure \eqref{eq:brag} will prove useful in describing spiral density waves in the presence of a highly turbulent magnetic-field configuration, for which KMHD becomes intractable.

\section{Linear Collisionless Shearing Waves}\label{sec:lw}

In this section, we solve for the propagation, damping, and---in the limit of strong scattering---turbulent excitation of spiral density waves in collisionless, differentially rotating plasmas. In all setups that we present, the waves studied possess no $z$ wavenumber. This is motivated by the fact that line-of-sight emission from waves with $k_z=0$ is stronger than those with a significant $k_z$; for example, \citet{kunz16} found in their shearing-box study of the collisionless MRI that non-axisymmetric density waves only appear after integrating over the $\ez$ direction (and then only weakly still). Additionally, when considering waves in a turbulent background magnetic field (\S\ref{sec:turbb}), we spatially average over the $z$ coordinate, so $k_z\neq 0$ waves are eliminated outright. Beyond neglecting $k_z$, we adopt the standard non-modal approach that describes non-axisymmetric, linear perturbations to the background state as an initial value problem for ``shearing waves''~\citep{glb65,johnson07}. In this approach, fluctuations in a plasma quantity $\chi$ are assumed to take the form
\begin{equation}\label{eq:form}
    \chi(t,\bb{r}) = \chi_0 + \delta \chi(t)\,\rme^{\imag k_x(t)x+\imag k_yy},
\end{equation}
where the radial wavenumber $k_x(t) = k_{x,0} - 2A_0 k_y t$ is time-dependent on account of the background flow shear. The primary advantage of this approach is that the spatial dependence of the parallel acceleration in \eqref{eq:kkin} and \eqref{eq:kmom} that is caused by the flow shear is exchanged in favor of a time dependence, reducing the governing equations to a system of time-dependent linear ODEs.

Given this approach, we consider four distinct scenarios. First, in \S\ref{sec:azimb} the background magnetic field is considered to be azimuthal and constant; this situation is amenable to an analysis using the KMHD set of equations, both when $\nu_{\rm eff}=0$ (scenario 1, \S\ref{sec:nu0}) and when $\nu_{\rm eff} \neq 0$ (scenario 2, \S\ref{sec:nun0}). In \S\ref{sec:azimb}, we investigate the impact of damping mechanisms on only the propagation of shearing waves as they transition from leading ($k_x(t)/k_y < 0$) to trailing ($k_x(t)/k_y > 0$); the actual excitation of spiral waves is not studied. The excitation process H\&P described requires a turbulent background state to supply (pseudo-potential) vorticity, which can eventually be converted into spiral waves. Being absent of turbulence and therefore vorticity, the laminar, $\bb{B}_0 = B_0\ey$ background state of \S\ref{sec:azimb} does not excite spiral density waves on its own. Therefore, in \S\ref{sec:turbb} we consider a turbulent background so that the spiral-wave excitation process can be studied as well. Because the KMHD equations become prohibitively difficult with which to work in the presence of a turbulent magnetic field, we only treat the limit of strong micro-instability scattering in \S\ref{sec:turbb} so that the simplifying approximation of Braginskii-MHD may be employed. With Braginskii-MHD, we first investigate the evolution of initially leading density waves similarly to scenarios 1 and 2 (scenario 3, \S\ref{sec:awaw}), allowing us to study the effects of damping alone. Once damping in the turbulent background field is well understood, we study initially leading vorticity fluctuations (scenario 4, \S\ref{sec:vwaw}), which are converted into spiral density waves and therefore serve as a combined study of damping and excitation. Altogether then, this work presents a study of the damping of collisionless spiral density waves in two extremes: laminar magnetic fields and isotropically tangled magnetic fields.

For ease of notation, we henceforth drop all ``0'' subscripts on the local rotation and shear frequencies, $\Omega_0\rightarrow\Omega$ and $A_0\rightarrow A$.

\subsection{Azimuthal Background Magnetic Field}\label{sec:azimb}

Beginning with the laminar azimuthal background magnetic field $\bb{B}_0 = B_0\ey$, we linearize the KMHD system (\ref{eq:kkin})--(\ref{eq:moments}) about the equilibrium shear flow $\bb{u}_0 = 2Ax\ey$ and an isotropic Maxwellian background distribution, $f(t,\bb{r},w_\perp,w_\|) = F_{\rm M,0}(w) + \delta f(t,\bb{r},w_\perp,w_\|)$, and insert the assumption of shearing waveforms \eqref{eq:form}.\footnote{It is likely that the actual distribution function $f$ in a collisionless accretion flow is not a small perturbation away from a Maxwellian. However, the regulation of temperature anisotropy by small-scale instabilities limits the ratio $T_\perp/T_\|$ to be very nearly unity on average~\citep{kunz16}. For these reasons, we make the simplifying assumption that the background distribution is an isotropic Maxwellian.} In the absence of scattering, the resultant system of equations is:
\begin{subequations}\label{eq:lin}
\begin{equation}\label{eq:klin}
    \biggl(\pD{t}{} + \imag k_y w_\|\biggr) \delta f + \Biggl[2\imag\biggl(\frac{w_\|}{v_{\rm th,i}}\biggr)^2k_y\delta u_{y} + \imag\biggl(\frac{w_\perp}{v_{\rm th,i}}\biggr)^2(k_x-2Ak_y t)\delta u_{x}
    - \imag k_yw_\|\frac{\delta p_{\|}}{p_0} + 2A \frac{\delta B_{x}}{B_0}\biggl(\frac{2w_\|^2-w_\perp^2}{v_{\rm th,i}^2}\biggr)\Biggr] F_{\rm M,0} = 0,
\end{equation}
\begin{equation}
    \D{t}{\delta u_{x}} = -\imag(k_x-2Ak_y t)\biggl(\frac{\delta p_{\perp}}{\rho_0}+\frac{B_0\delta B_{y}}{4\pi\rho_0}\biggr)+\imag k_y\frac{B_0\delta B_{x}}{4\pi\rho_0}+2\Omega \delta u_{y},
\end{equation}
\begin{equation}
    \D{t}{\delta u_{y}} = -\imag k_y\frac{\delta p_{\|}}{\rho_0}-2(\Omega+A)\delta u_{x},
\end{equation}
\begin{equation}
    \D{t}{\delta B_{x}} = \imag k_y B_0 \delta u_{x},
\end{equation}
\begin{equation}\label{eqn:lindivB}
    k_y\delta B_{y} = -(k_x-2Ak_y t)\delta B_{x}.
\end{equation}
\end{subequations}
In \eqref{eqn:lindivB}, we have used the solenoidality condition on the perturbed magnetic field, rather than the induction equation, to obtain the $y$ component of the magnetic field. 

In all solutions obtained while employing this azimuthal $\bb{B}_0$ configuration, we use the same initial conditions for the spiral wave: a leading perturbation is initialized to $u_x$ alone, with $k_{x,0} = -10k_y$. This excites a fast magnetosonic wave (the MHD/KMHD analog of an acoustic wave), which we evolve until the radial wavenumber is exactly reversed ($k_x = 10k_y$).\footnote{In the collisionless limit, the $k_z=0$ shearing fast wave is entirely decoupled from the non-axisymmetric MRI~\citep{bh92_p4}, so we need not worry about instability affecting our investigation of collisionless damping. In ideal MHD, however, \citet{mamatsashvili2026} have shown the two waves to be coupled, and swing-amplified MRI fluctuations at least partially convert into spiral density waves for laminar magnetic-field configurations.} Although we vary other parameters, like the ratios $\Omega/k_yv_{\rm th,i}$ and $\beta$, we do not modify the initial wavenumber ratio.

When scattering effects are included, the density and temperature in \eqref{eq:coll} are perturbed to obtain the linearized Krook operator:
\begin{equation}\label{eq:lincoll}
    \left( \frac{\partial \delta f}{\partial t}\right)_{\rm scatt,eff} \approx -\nu_{\rm eff}\left[ \delta f(t,w_\perp,w_\|) - F_{\rm M,0}(w)\left( \frac{\delta \rho}{\rho_0} - \frac{3}{2}\frac{\delta T}{T_0} + \frac{w^2}{v_{\rm th,i}^2}\frac{\delta T}{T_0}\right) \right],
\end{equation}
which is appended to the right-hand side of \eqref{eq:klin} when necessary. Although the above system is significantly simplified from \eqref{eq:kkin}, it is still not analytically tractable in general. Therefore, we pursue numerical solutions to \eqref{eq:lin} as initial value problems using a Hermite--Laguerre polynomial decomposition technique (see Appendix A of \citealt{mks23} for a description of this method as applied to the non-shearing-sheet KMHD equations).

\subsubsection{Solutions for $\nu_{\rm eff} = 0$}\label{sec:nu0}

We first present solutions to the system \eqref{eq:lin} without any scattering ($\nu_{\rm eff}=0$).
\begin{figure}
\centering
\includegraphics[width=0.90\linewidth]{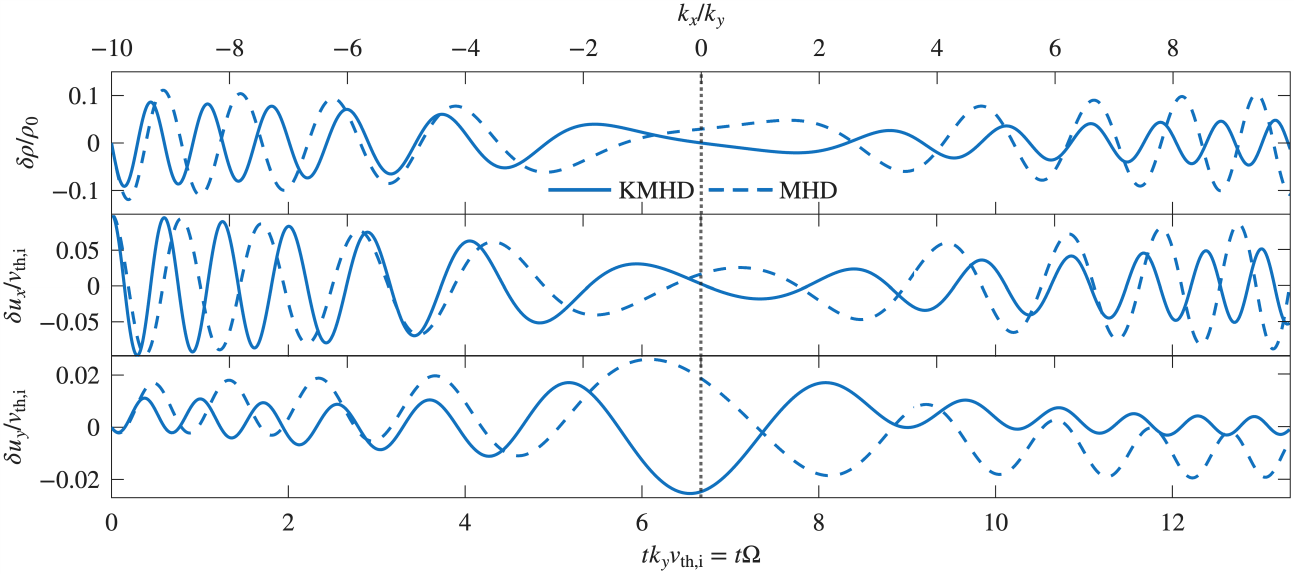}
\caption{Evolution of KMHD (solid) and ideal MHD (dashed) spiral density waves in an azimuthal magnetic field with $\beta=5$, initialized as leading $\delta u_x = 0.1v_{\rm th,i}$ perturbations. The vertical dotted line denotes the swing time when $k_x(t)=0$. Moderate damping due to resonant wave--particle interaction is reflected in the fact that at late times the KMHD wave amplitude is noticeably lower than that of the MHD wave.}
\label{fig:exam}
\end{figure}
To orient the reader, in figure \ref{fig:exam} we present the (numerically solved) evolution of an example collisionless leading fast wave (solid lines) with $k_y = \Omega/v_{\rm th,i}$ in a plasma with $\beta=5$ (here, $\beta=\beta_{\perp,0}=\beta_{\|,0}$ is that of the uniform background state). The density fluctuation (top panel) behaves in a manner that is largely typical of shearing density waves: being excited at $|k_x| \gg k_y$, its frequency $\omega$, which is proportional to $|k|$, decreases over time as $|k_x|$ decreases from the shear. As it does so its amplitude decreases, a consequence of (near) conservation of wave action preserving the ratio $E/\omega$~\citep[with $E$ the energy of the wave perturbation;][]{johnson07,hp09_1}. The amplitude reaches a minimum at the swing time when $k_x=0$, which is marked by the vertical dotted line, and then begins to grow again from wave-action conservation. The behavior of $\delta u_x$ and $\delta u_y$ is also predictable: at early times when $|k_x|\gg k_y$, the flow perturbation is dominated by its radial component $\delta u_x$, although it exchanges energy with the azimuthal perturbation $\delta u_y$ when $|k_x| \ll k_y$. We also solve the exact same initial value problem for an MHD plasma (dashed lines). Much of the behavior is similar, however the wave frequencies are slightly different and, more importantly, the MHD wave has a noticeably larger amplitude than the KMHD wave at the end of its evolution. This difference results from the collisionless damping in KMHD that is absent from ideal MHD, which, when present, also breaks perfect wave action conservation in the collisionless wave. Seeing as the collisionless wave amplitude appears to grow from action conservation at a similar rate to the MHD wave long after the swing, it seems that most of the damping must have occurred when $|k_x| \lesssim k_y$. 

The example shown in figure~\ref{fig:exam}, although illustrative, does not reveal fully the underlying physics of how collisionless damping operates in these waves. Furthermore, it is valuable to generalize these results to a range of $\beta$ and $\Omega/k_yv_{\rm th,i}$. This generalization can be accomplished both by an (approximate) analytic description, and by a survey of numerical solutions with a similar setup but varied parameters. In what follows, we take both approaches to explore the problem more thoroughly, beginning with the former. 

The analytic approach we adopt is a lowest-order WKB approximation, where we solve for the damping of these collisionless fast waves in the limit of slow shear compared to the field-parallel sound-crossing time, i.e., $|A|\ll k_y v_{\rm th,i}$. The WKB approximation is frequently used to describe sheared waves in accretion flows, especially the non-axisymmetric MRI~\citep[e.g.,][]{bh92_p4,ko01,jg05,bh06}. For the MRI, however, the common choice for a WKB parameter is $k_y/|k_x| \ll 1$. Because the damping of spiral waves appears strongest when $|k_x| \lesssim k_y$, we must instead use the low-shear approximation (similarly to H\&P). As is typical of lowest-order WKB approximations, our steps are to first solve for the instantaneous (complex) frequency of these waves given a constant $k_x$, then restore the time dependence of $k_x(t)$ and integrate the wave eikonal forward in time to obtain the evolution of the wave. Because we are concerned with damping, our focus will be on the complex part of the frequency. Without further ado, we now neglect $A$ to leading order in the linearized system \eqref{eq:lin}, and substitute perturbations of the form $\delta \chi(t) = \delta\chi\exp(-\imag\omega t)$. After taking the appropriate moments of $\delta f$, the resultant system can be solved for the instantaneous dispersion relation:
\begin{equation}\label{eq:disp}
    \biggl\{\zeta^2-\frac{1}{\beta}\biggl(1+\frac{k_x^2}{k_y^2}\biggr) + \frac{k_x^2}{k_y^2} \,\zeta Z(\zeta) + 8\zeta^2\biggl(\frac{\Omega}{k_yv_{\rm th,i}}\biggr)^2\bigl[1+\zeta Z(\zeta)\bigr]\biggr\}\biggl(\zeta^2-\frac{1}{\beta}\biggr) \equiv \epsilon(\zeta,\bb{k}) = 0.
\end{equation}
Here, $\zeta \equiv\omega/(k_yv_{\rm th,i})$ and $Z(\zeta) \equiv \pi^{-1/2}\int_{-\infty}^{\infty}\rmd x\, \exp(-x^2)/(x-\zeta)$ is the ``plasma dispersion function.'' The Alfv\'en-wave solution, given by $\zeta = \pm \beta^{-1/2}$ in the right-most parentheses of \eqref{eq:disp}, decouples from the compressive fluctuations (in curly brackets) given our assumptions. With $\Omega\sim |A|$ in a Keplerian flow, one could also drop the $\Omega$ term in \eqref{eq:disp}; for the sake of generality, however, we retain it for now. Because of the non-analytic nature of the plasma dispersion function, we can only obtain approximations to the instantaneous frequency of these fast waves. Given that only a small amount of damping occurs in figure \ref{fig:exam} over many periods of the wave, let us then adopt the simplifying assumption $\zeta_{r} \gg \zeta_{i}$ (with the subscripts ``$r$'' and ``$i$'' denoting real and imaginary parts). Expanding the dispersion relation about $\zeta \approx \zeta_{r}$ leads to
\begin{equation}\label{eq:asyma}
    \epsilon(\zeta,\bb{k}) \approx \epsilon(\zeta_{r},\bb{k}) + \imag \zeta_{i}\pD{\zeta}{}\epsilon(\zeta_{r},\bb{k}) \approx 0.
\end{equation}
The assumption that $\zeta_{i} \ll \zeta_{r}$ is consistent with taking $\epsilon_{i} \ll \epsilon_{r}$, so that the real and imaginary parts of this equation become
\begin{equation}\label{eq:asymp}
    \epsilon_{r}(\zeta_{r},\bb{k}) \approx 0
    \, \qquad \mathrm{and} \qquad \zeta_{i} \approx -\frac{\epsilon_{i}(\zeta_{r},\bb{k})}{\partial_{\zeta_{r}}\epsilon_{r}(\zeta_{r},\bb{k})}.
\end{equation}
$Z(\zeta)$ is still non-analytic, so we expand in the limit of $|\zeta_{r}| \gg 1$ (the validity conditions for which we determine {\em a posteriori}) and solve for the wave frequency to obtain
\begin{equation}\label{eq:zr}
    \zeta_{r} \approx\pm \sqrt{\frac{1}{\beta}\left(1+\frac{k_x^2}{k_y^2}\right)+\frac{k_x^2}{k_y^2}+\frac{4\Omega^2}{k_y^2v_{\rm th,i}^2}}.
\end{equation}
To determine when \eqref{eq:zr} is valid, the minimum value of $|\zeta_{r}|$ occurs when $k_x = 0$, giving $|\zeta_{r}| \geq \sqrt{\beta^{-1}+4\Omega^2/k_y^2v_{\rm th,i}^2}$. Given that we have already assumed $|A|\sim \Omega \ll k_y\vth$, the expansion of $Z(\zeta)$ that we have employed is acceptable when $\beta \lesssim 1$. We now insert equation~\eqref{eq:zr} into the second equation of \eqref{eq:asymp} to obtain
\begin{equation}\label{eq:damp}
    \zeta_{i} = -\frac{\sqrt{\pi}}{2}\biggl\{ \frac{|k_x|}{k_y} + \frac{8\Omega^2}{k_y^2v_{\rm th,i}^2}\biggl[ \frac{1}{\beta}+\frac{|k_x|}{k_y}\biggl(1+\frac{1}{\beta}\biggr)+ \frac{4\Omega^2}{k_y^2v_{\rm th,i}^2}\biggr] \biggr\}\exp \biggl[ -\frac{1}{\beta} - \frac{k_x^2}{k_y^2}\biggl(1+\frac{1}{\beta}\biggr)-\frac{4\Omega^2}{k_y^2v_{\rm th,i}^2}\biggr].
\end{equation}
Because all terms in the exponential are negative, $\zeta_{i}$ is largest when those terms are most nearly zero. Given a $\beta$ and $\Omega/k_yv_{\rm th,i}$, the strongest damping therefore occurs when $|k_x|/k_y \ll 1$, and the wavevector is most nearly aligned with the azimuthal background field. Physically, this is because the ponderomotive force that drives Landau/Barnes damping also becomes aligned with the field direction when $k_x=0$, and so the resonant ions, which can only move freely along $\bb{B}$, are able to respond to that force maximally. As $|k_x|$ grows however, the force misaligns with $\bb{B}$, and the exponential term rapidly becomes small, hence why there are no obvious signs of damping at early or late times in figure~\ref{fig:exam}. Additionally, although \eqref{eq:damp} is only valid at moderate to low $\beta$, we can again see from the exponential that moderate $\beta$ leads to more damping than very low $\beta$ (due to the $-\beta^{-1}$ term). This can be attributed to the fact that thermal fluctuations contribute more to the wave energy at $\beta \sim 1$ than they do at $\beta\ll 1$, so any damping of those thermal fluctuations by resonances in the distribution function will have a more significant impact on the wave at larger $\beta$. In fact, although \eqref{eq:damp} is not valid in the limit $\beta \gg 1$, the previous statement still implies that once the magnetic field contributes a negligible amount to the wave energy, $\zeta_{i}$ should become effectively independent of $\beta$ because only the thermal contributions will matter. In that limit, the damping rate must instead be found by dropping the term $\propto \beta^{-1}$ in the curly braces of \eqref{eq:disp}, and numerically solving for $\zeta$, which will be a function of only $\Omega/k_yv_{\rm th,i}$ and $k_x/k_y$ (or, considering the Keplerian WKB limit, $k_x/k_y$ alone).

With an expression for $\zeta_{i}(k_x)$ in hand, the final step in estimating the role of damping in our current spiral wave setup is to integrate the eikonal $-\zeta_{i}(k_x(t))k_yv_{\rm th,i}\rmd t$ forward in time as is typical of the WKB approximation. Using the shearing wavenumber relationship $k_x(t) = k_{x,0}-2Ak_yt$, this time integral can be exchanged in favor of an integral over $k_x$, which we provide here for a symmetric interval about $k_x=0$:
\begin{equation}\label{eq:int}
    \delta \rho(t{=}k_{x,0}/Ak_y) = \delta \rho(t{=}0) \exp \left[ \frac{1}{2Ak_y}\int\limits_{k_{x,0}}^{-k_{x,0}} \rmd k_x \, \zeta_{i}(k_x)k_yv_{\rm th,i} \right].
\end{equation}
We have given \eqref{eq:int} for a symmetric interval about $k_x=0$ because it would not be strictly correct otherwise: as we alluded to in the discussion of figure~\ref{fig:exam}, wave-action conservation causes the amplitude of a spiral density wave to change in time in a manner that is symmetric about the swing time. Because wave-action conservation does not lead to damping, it does not contribute to the suppression of spiral density waves. We therefore remove its effect by considering how the amplitude of our spiral density wave changes between $k_x(t) = k_{x,0}$ and $-k_{x,0}$.\footnote{For a detailed discussion of the role of wave-action conservation in sheared MHD waves, see \citet{johnson07}.} We lastly note that, because \eqref{eq:damp} has a Gaussian form in $k_x$, the integral \eqref{eq:int} can technically be evaluated analytically. For finite $k_{x,0}/k_y$ however, its form is much too cumbersome to include explicitly, so we leave the reader to evaluate it analytically or numerically as they deem necessary. Regardless, the dependence of the cumulative damping on $\beta$ and $k_x/k_y$ remains similar to that found in the discussion of \eqref{eq:damp}.

\begin{figure}
\centering
\includegraphics[width=0.72\linewidth]{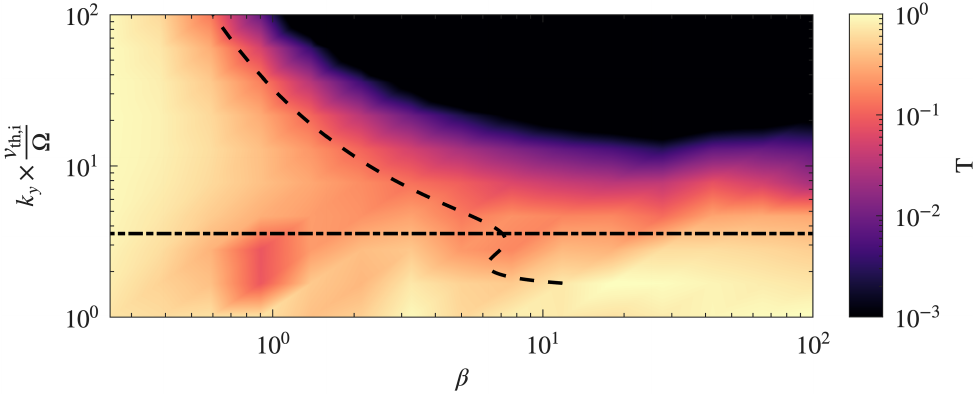}
\caption{Efficiency of transfer from leading to trailing for fast waves with $\nu_{\rm eff}=0$, as a function of $\beta$ and $k_y$. Damping is most effective for large-$k_y$ waves because they experience a larger number of wave periods for a given time over which $k_x(t) \rightarrow -k_{x,0}$. The dashed and dash-dotted lines are WKB predictions of the $\rm T=0.1$ curves in the low-to-moderate $\beta$ and high-$\beta$ limits, respectively. The lack of damping at low $k_y$ indicates that Landau/Barnes damping may not significantly affect spiral waves excited by the process described in H\&P.}
\label{fig:nu0}
\end{figure}
Having discussed the damping of the collisionless fast wave analytically, we now perform a survey of numerical solutions to this problem over $\beta$ and $\Omega$ to see these effects in practice. In figure \ref{fig:nu0}, we integrate the full set of equations \eqref{eq:lin} numerically to calculate the amount of damping that is incurred by fast waves over a range of $\beta$ and $k_yv_{\rm th,i}/\Omega$ for a period of time that is symmetric about the swing time ($k_x = -10 k_y \rightarrow 10k_y$). Dissipation is measured in terms of the following quantity:
\begin{equation}\label{eq:transc}
    {\rm T} \equiv \frac{|\delta \rho_{\textrm{p-p}}(t{=}k_{x,0}/Ak_y)|}{|\delta \rho_{\textrm{p-p}}(t{=}0)|},
\end{equation}
which measures the fraction of the density wave amplitude that is successfully transferred from leading to trailing. Here, the subscript ``p-p'' refers to the fact that we use the peak-to-peak amplitudes of the waves at the beginning and end of the numerical solution to determine $\rm T$.\footnote{This measure for $\rm T$ motivates the choice $k_x = -10 k_y \rightarrow 10k_y$ for the evolution of our fast mode. A shorter evolution interval means relatively few wave periods before and after the swing from which to calculate the peak-to-peak amplitude.} As mentioned earlier, with the wave evolved for a period centered on the swing time, the effects of action conservation on the amplitude are irrelevant. The most clear trend in figure~\ref{fig:nu0} is that large-$k_y$ (or alternatively small-$\Omega$) waves are the ones that are most effectively damped. This comes about because the number of wave periods they experience during a swing (when damping is strongest) is larger than for a wave with small $k_y$, so, given the same ratio $|\zeta_{i}|/\zeta_{r}$, more of the fast-wave energy can be lost during the swing to Landau/Barnes damping. By comparison, this also means that lower-$k_y$ waves, which swing quickly compared to their oscillation, are more weakly damped. As anticipated from Equation~\eqref{eq:damp}, as well as our discussion of the alternative $\beta \gg 1$ limit, damping also becomes more effective as $\beta$ increases from $\ll 1$ to $\sim 1$, eventually steadying once $\beta \gg 1$. To compare to our WKB approximation, we include a dashed line representing the prediction for $\rm T=0.1$ that is obtained by evaluating the integral \eqref{eq:int} given the damping rate \eqref{eq:damp}. At low, and surprisingly even moderate $\beta$, this estimate yields a reasonable approximation to the actual $\rm T=0.1$ curve from the numerical simulations. We also evaluate the same integral \eqref{eq:int} for a numerical solution to \eqref{eq:disp} that neglects the $\beta^{-1}$ term (a high-$\beta$ approximation). The prediction that it yields for $\rm T=0.1$ (the dash-dotted line) is a good approximation to the numerical solutions of \eqref{eq:lin} in the high-$\beta$ limit.

Perhaps the most important finding that we have obtained in this section is that Landau/Barnes damping is inefficient at damping spiral waves with moderate-to-low values of $k_y v_{\rm th,i}/\Omega$, a consequence of their relatively brief time spent with $\bb{k}\, \| \,\bb{B}_0$. This is especially notable given that H\&P found the spiral-wave excitation process in MHD to be most effective for waves having $k_y v_{\rm th,i}/\Omega \sim 1$. Accordingly, in an accretion flow that is free of particle scattering and that possesses a relatively uniform, dominantly azimuthal field, Landau/Barnes damping is not expected to damp spiral waves significantly. We now test whether this conclusion still holds in the presence of particle scattering. 

\subsubsection{Solutions for Wave-driven $\nu_{\rm eff}$}\label{sec:nun0}

In the presence of sufficient scattering ($\nu_{\rm eff} \gtrsim |\eb\bcdot \bb{k}|\,v_{\rm th,i}$), the resonances that drive Landau/Barnes damping are interrupted and the dominant damping mechanism becomes viscous damping. As a fast wave propagates, it generates pressure anisotropy through the approximate conservation of the particle adiabatic invariants~\citep{cgl56}. Scattering, on the other hand, isotropizes the distribution, and therefore the pressure tensor, effectively thermalizing the portion of the wave energy responsible for producing that pressure anisotropy and viscously damping the wave. Unlike Landau/Barnes damping, viscous damping can occur regardless of the wavevector's alignment with the magnetic field. It is therefore not limited to operating near the swing time, but rather can occur even when the wave is leading or trailing. Considering this extended range of wavenumbers over which viscous damping occurs then, a spiral density wave that is subject to weak scattering may experience more damping than one that is entirely free of scattering (or alternatively, one in which scattering is so strong that LTE is enforced at all times, e.g., the MHD solution of figure~\ref{fig:exam}).

As discussed in \S\ref{sec:nlte}, the accretion-flow environments that we investigate here are practically free of Coulomb collisions. Therefore, any scattering that may be present is instead expected to be a result of pressure-anisotropy-triggered micro-instabilities (e.g., the mirror and firehose). The unperturbed background state that we study in this section, namely $\bb{B}_0 = B_0\ey$ and $\bb{u}_0 = 2Ax\ey$, does not produce any pressure anisotropy on its own because $\eb_0\eb_0\bdbldot\grad\bb{u}_0=0$. However, fast magnetosonic waves do produce a fluctuating pressure anisotropy as they propagate, and at sufficiently high $\beta$, that pressure anisotropy can become large enough in magnitude to surpass the mirror and firehose instability thresholds \eqref{eq:thresh}. In \citet{mks23}, the authors found that a fast wave propagating perpendicular to a uniform magnetic field produces enough $\Delta p$ to trigger the mirror and firehose instabilities when it perturbs the magnetic field such that $|\delta B|/B_0 \gtrsim 2/\beta$. Similarly, high-$\beta$ fast waves propagating parallel to the magnetic field (also known as ion-acoustic waves) trigger micro-instabilities whenever $|\delta \rho|/\rho_0 \gtrsim 2/\beta$~\citep{kunz20}. At $\beta \gtrsim 10$, fast waves with supercritical amplitudes can still be reasonably treated according to linear theory, because the timescales associated with nonlinear processes, such as wave steepening, remain larger than the linear wave period. As we saw in figure~\ref{fig:exam}, shearing-wave amplitudes are not constant in time, so a generic amplitude limit cannot be as easily defined here. Nonetheless, we can still investigate how self-generated micro-instabilities and their scattering affect the propagation and damping of spiral density waves of various amplitudes in the azimuthal-$\bb{B}_0$ setup employed here.

\begin{figure}
\centering
\includegraphics[width=0.93\linewidth]{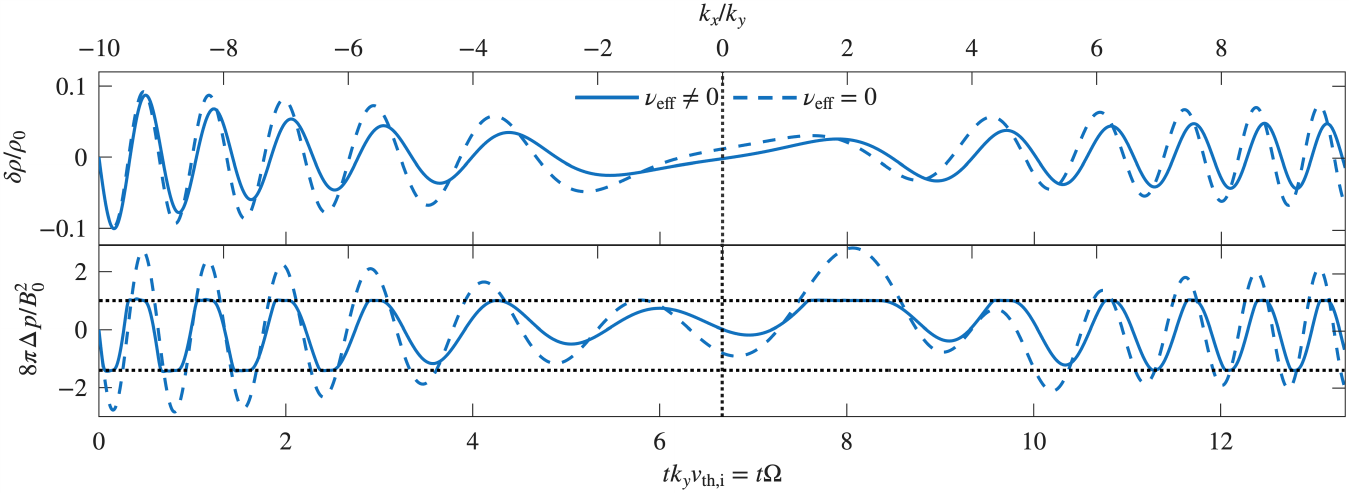}
\caption{Evolution of spiral density wave $\delta \rho$ (top) and $\Delta p$ (bottom) with similar initial and background conditions to \S\ref{sec:nu0}, but with (solid) and without (dashed) scattering from self-generated micro-instabilities. Here, $\beta=30$ to ensure that a wave of modest amplitude can reach the instability thresholds \eqref{eq:thresh}, denoted by horizontal dotted lines on the bottom panel. A moderate increase in damping results from the thermalization of wave energy due to micro-instability scattering.}
\label{fig:exam_nu}
\end{figure}
In figure~\ref{fig:exam_nu}, we show the evolution of $\delta \rho$ (top) and $\Delta p$ (bottom) in an example spiral fast wave that is initialized in the same manner as in \S\ref{sec:nu0} (a $u_x$ perturbation alone, with $k_{x,0} = -10k_y$ and $\Omega = k_yv_{\rm th,i}$) but which is subject to effective scattering from self-generated micro-instabilities (solid lines). The background plasma $\beta = 30$, which was chosen to ensure that the wave amplitude ($\delta u_{x,0} = 0.1v_{\rm th,i}$) would be more than large enough to reach both instability thresholds (marked by the horizontal dotted lines on the bottom panel). To mimic the nonlinear phase of the mirror and firehose instabilities, whenever the pressure anisotropy exceeds the thresholds \eqref{eq:thresh}, we activate a nonzero $\nu_{\rm eff}$. The $\nu_{\rm eff}$ that we use is calculated directly from the linearized form of \eqref{eq:scat}, which (when active) varies in time according to the wave amplitude and vector and is given by
\begin{equation}\label{eq:linscat}
    \nu_{\rm eff}(t) = \beta_0 \left| 6A\frac{\delta B_x(t)}{B_0}+2\imag k_y \delta u_y(t)-\imag k_x(t) \delta u_x(t) \right|.
\end{equation}
To highlight the role of scattering by comparison, we also show the evolution of an identical wave, but without any scattering from micro-instabilities (dashed lines). The most obvious difference between the two waves is that the pressure anisotropy is not permitted to exceed the instability thresholds when micro-instability scattering is included, while it is allowed to exceed the thresholds freely when micro-instability scattering is excluded. This demonstrates that some of the wave energy is indeed being thermalized when $\nu_{\rm eff}$ is active. The consequences of that thermalization can be seen in the slight damping of $\delta \rho$ (as compared to the $\nu_{\rm eff} = 0$ wave). Notably, viscous damping due to micro-instability scattering begins immediately, unlike in figure \ref{fig:exam} where appreciable damping occurs only near $k_x=0$. This agrees with the expectation that viscous damping can occur even when the wavevector is not aligned with the magnetic-field direction. That said, the pressure anisotropy is not driven unstable when $|k_x| \lesssim k_y$---a consequence of action conservation lowering the amplitude of the wave near the swing time. Therefore, in the spiral fast wave shown here, micro-instability scattering is not driven during the same period when Landau/Barnes damping is most effective, allowing the two effects to supplement each other (as opposed to $\nu_{\rm eff}$ exclusively suppressing Landau/Barnes damping). This result is, however, dependent on the initial wave amplitude and $\beta$; an increase in either of those quantities may cause the wave $\Delta p$ to reach the thresholds even at low $|k_x|$. To explore this possibility, we now study the problem for varying $\beta$ and wave amplitudes.

\begin{figure}
\centering
\includegraphics[width=0.84\linewidth]{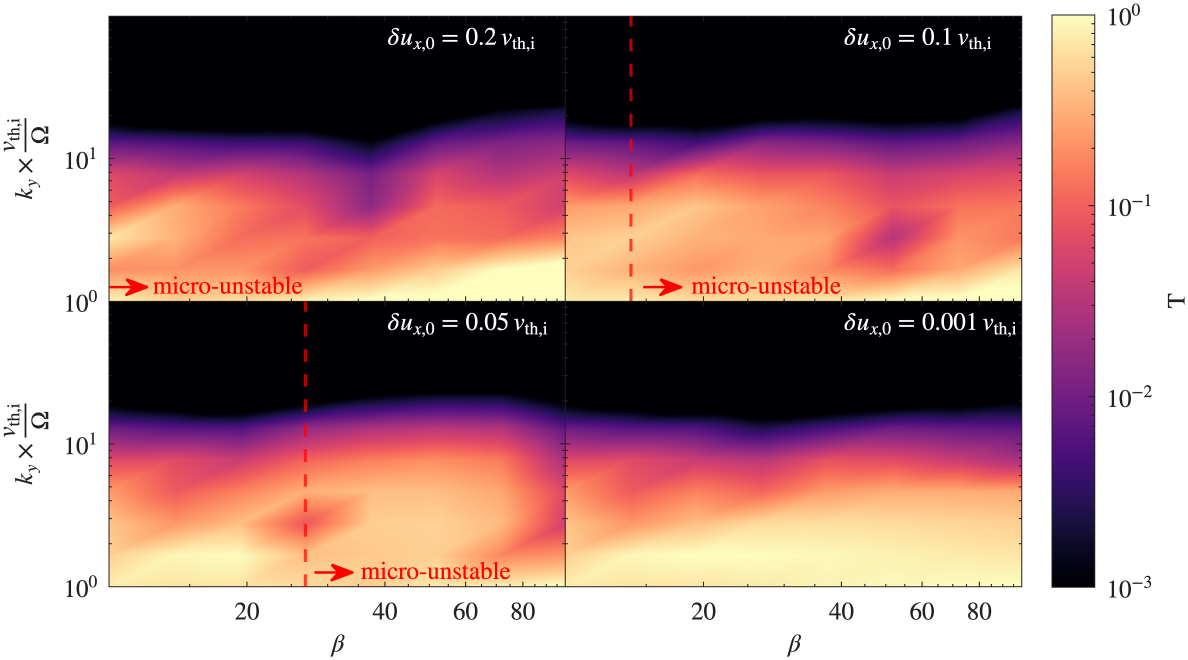}
\caption{Efficiency of transfer from leading to trailing for fast waves with self-generated scattering, as a function of $\beta$ and $k_y$, for four different initial $\delta u_{x}$ amplitudes. Red vertical dashed lines with arrows indicate the regions in which the given amplitude/$\beta$ combination is sufficient to trigger instability scattering. For $\delta u_{x,0} = 0.2 v_{\rm th,i}$ all parameters lead to the triggering of instabilities, whereas for $\delta u_{x,0} = 0.001 v_{\rm th,i}$, none were sufficient given the shown range of $\beta$, making it equivalent to figure \ref{fig:nu0}. For the parameter regime studied here, self-generated mirror/firehose scattering leads to a subtle increase in the damping of spiral density waves, which grows with $\delta u_{x,0}$.}
\label{fig:scan_nu}
\end{figure}
In figure \ref{fig:scan_nu}, we calculate the same transfer coefficient $\rm T$ as defined in \eqref{eq:transc}, but for waves of different initial amplitudes (otherwise given the same initial conditions), including viscous damping from self-generated micro-instability scattering. Here, the scan in $\beta$ is only performed for $\beta\geq 10$ because waves that are sufficiently large to trigger instabilities at $\beta < 10$ are not described well by linear theory. Vertical dashed red lines (with accompanying arrows) denote the regions of parameter space in which the wave amplitude and plasma $\beta$ are sufficiently large to trigger micro-instabilities. In the top left panel, instabilities are triggered everywhere, while in the bottom right, they are triggered nowhere (hence, the bottom right panel yields the same values for $\rm T$ as figure \ref{fig:nu0}). As a baseline against which to compare this survey, in \citet{hp09_2} the authors studied spiral-wave excitation from MRI turbulence directly using MHD simulations and found that wave amplitudes peaked around $\delta u_x \sim 0.1 v_{\rm th,i}$. The overall trend we observe is that anomalous scattering from wave-generated micro-instabilities increases viscous damping with increasing wave amplitude. The increase in damping is, however, subtle, and does not lead to a strict decrease in $\rm T$ at \textit{all} $\beta$ and $k_y$. The notable exception is the very high-$\beta$, low-$k_y$ regime at $\delta u_{x,0} = 0.2v_{\rm th,i}$; there, the combination of large amplitude and high $\beta$ has allowed the wave to reach the strong-scattering regime ($\nu_{\rm eff} \gtrsim 100 k_yv_{\rm th,i} \gg |\eb\bcdot \bb{k}|\,v_{\rm th,i}$), in which case its dynamics approximate that of ideal MHD because the effective mean free path has become much shorter than the wavelength (see \citet{kunz20} and \citet{mks23} for similar examples of the strong-scattering limit). Furthermore, because the wave possesses low $k_y$, it is sheared (and therefore damped) for fewer wave periods than a comparable wave at larger $k_y$. Accordingly, the rate of damping has decreased to near zero and $\rm T\rightarrow 1$.~\footnote{At even higher values of $\beta$,it may be that the plasma behaves close enough to ideal MHD that the coupling between non-axisymmetric MRI and spiral density waves is recovered, as in \citet{mamatsashvili2026}. We leave such a possibility to future work.}

In summary, self-generated micro-instability scattering leads to viscous damping that increases with the amplitude of the spiral density wave; however, the increase in damping is subtle, and when wave amplitudes approach nonlinear values ($\delta u_{x,0} \sim 0.2v_{\rm th,i}$) this trend reverses for $\beta\gtrsim 50$ as the plasma mimics a strongly collisional fluid.

\subsection{Turbulent Background Magnetic Field}\label{sec:turbb}

We have so far studied the damping of spiral density waves in a uniform, azimuthal magnetic field, including when scattering from pressure-anisotropy-driven micro-instabilities is absent or active. Because there is no turbulence in this laminar background state, and therefore no vorticity, these studies allow us to probe only the damping and propagation of spiral density waves, not their excitation. Let us now consider a situation in which turbulence is also present in the shearing sheet, which allows us to address the propagation, damping, \textit{and} excitation of spiral density waves. The requisite turbulence is assumed to be driven by the magnetorotational instability and/or the closely related magnetoviscous instability (unique to collisionless and weakly collisional differential flows; \citealt{qdh02,balbus04}). This turbulent state implies that there is also a near-volume-filling, nonzero $\nu_{\rm eff}$ that results from micro-instabilities being driven by the turbulent pressure anisotropy. We now develop a set of equations to describe spiral waves in this turbulent system.

Unfortunately, deriving a numerically integrable, linearized KMHD system for spiral waves in a highly turbulent magnetic field is too onerous a task for this work. However, in the presence of sufficient scattering from micro-instabilities we may employ the Braginskii-MHD closure for $\Delta p$, which makes it easier to obtain a tractable system of equations. The use of Braginskii-MHD requires scattering that is strong enough to maintain $\Delta p \ll p$, which is normally satisfied by linear wave dynamics when $\nu_{\rm eff} \gg kv_{\rm th,i}$. Physically, this criterion states that a particle must be scattered several times before it transits one wavelength, in order to ensure that the distribution does not drift far from LTE. In our system, tangling of the magnetic field by the turbulence effectively increases the path length that a field-following particle must travel in order to traverse one wavelength of the actual spiral wave. Due to this increased path length, particles can experience more scattering per wavelength transit than they would otherwise in a uniform field, so we expect $\nu_{\rm eff} \gtrsim kv_{\rm th,i}$ to be a sufficient condition in practice. To determine $\nu_{\rm eff}$, equation~\eqref{eq:scat} requires an estimate for $|3\eb\eb\bdbldot\grad\bb{u}-\grad\bcdot\bb{u}|$ in this turbulent-field configuration. Because the magnetic field is no longer purely azimuthal, the sheared flow $\bb{u}_0 = 2Ax\ey$ now makes a nonzero contribution to $\eb\eb\bdbldot\grad\bb{u}$. Given that $|\eb|$=1, the root-mean-square value of $|3\eb\eb\bdbldot\grad\bb{u}_0|$ will be ${\sim}|A|$. We insert this root-mean-square estimate for the rate of strain into \eqref{eq:scat}, finding that scattering is driven at a rate $\nu_{\rm eff} \sim \beta|A|$ (${\sim} \beta\Omega$ for Keplerian rotation, neglecting order-unity factors).\footnote{The scattering rate $\nu_{\rm eff}$ is also perturbed by the wave-generated pressure anisotropy, as in \S\ref{sec:nun0}. However, the wave contribution to $\nu_{\rm eff}$ \eqref{eq:linscat} is proportional to the spiral-wave amplitude, which is small relative to that of the turbulence. Additionally, although these instabilities are not driven \textit{everywhere} in the domain at a given instant, \citet{arzamasskiy23} found that they intermittently occupy each part of the domain sufficiently well that Braginskii-MHD is valid on (volumetric) average.} Braginskii-MHD is therefore a valid assumption when $\beta \Omega \gtrsim kv_{\rm th,i}$, or, in terms of the scale height $H \equiv c_{\rm s}/\Omega$, $\beta \gtrsim kH$ (where $c_{\rm s}=\sqrt{5p/3\rho}$ is the adiabatic sound speed). Because spiral waves are excited near $k\sim k_y \sim H^{-1}$ (H\&P), $\beta \gtrsim kH$ translates into the requirement that $\beta \gtrsim 1$ shortly after excitation (when $|k_x| \lesssim |k_y|$), but it eventually requires $\beta \gtrsim |k_x/k_y|$ long after excitation (once $|k_x| \gg |k_y|$). 

Given the above assumptions, we may begin shaping the Braginskii-MHD equations into a linearized, solvable system. We first re-write the shearing-sheet momentum equation in a form that is more convenient with which to work:
\begin{equation}\label{eq:bmom}
    \left( \pD{t}{} + 2Ax\pD{y}{}\right)\rho\delta \bb{u}=-\grad p-2\Omega\ez\btimes \rho\delta \bb{u}-2A\rho \delta u_x\ey - \grad \bcdot \msb{T},
\end{equation}
where $\bb{u}_0$ has been separated out of the bulk flow, {\it viz}.~$\bb{u} = \delta \bb{u} + 2Ax\ey$. $\msb{T}$ is a tensor that captures all nonlinear contributions to the perturbed momentum $\rho \delta \bb{u}$, namely, the Reynolds, Maxwell, and viscous stresses:
\begin{equation}\label{eq:stress}
    \msb{T} = - \frac{\bb{BB}}{4\pi} + \msb{I}\frac{B^2}{8\pi} + \rho \delta \bb{u}\delta\bb{u} - \left( \eb\eb-\frac{1}{3}\msb{I}\right)\Delta p.
\end{equation}
Because we now have turbulence, spiral waves, and a sheared flow present in the shearing sheet, but seek only to model the spiral waves and their interaction with the sheared flow, in what follows it is prudent to differentiate turbulent quantities from wave and sheared-flow quantities. We distinguish their contributions to each fluid variable by introducing the subscripts ``ws'' and ``tu'', which are short for ``wave and sheared flow'' and ``turbulence''. In ``ws'' quantities, we group the spiral-wave and sheared-flow contributions to the fluid variables together according to a shared property: none of them depend on the $z$ coordinate. The uniform parts of the density and pressure are therefore also contained within $\rho_{\rm ws}$ and $p_{\rm ws}$, respectively. By contrast, ``tu'' quantities, which refer to fluctuations resulting from the turbulence, vary randomly with the $z$ coordinate. It is therefore implied that when fluid quantities are averaged over the $z$ coordinate, e.g., $\mnz{\bb{u}}$, that the turbulent component vanishes, leaving only the ``ws'' component ($\mnz{\bb{u}} = \bb{u}_{\rm ws}$). Furthermore, $\delta \bb{u}_{\rm ws} = \bb{u}_{\rm ws} - \bb{u}_0$ contains only the wave perturbation to the flow, unlike $\delta \bb{u}$, which contains the wave and turbulence contributions, but not the sheared flow. We make use of these properties in the calculations that follow.

With the ``ws'' and ``tu'' decomposition of the fluid properties, the contributions to the pressure anisotropy can be separated out as
\begin{equation}\label{eq:sepdp}
    \Delta p = \Delta p_{\rm tu} + \Delta p_{\rm ws} \approx \Delta p_{\rm tu} + 3\frac{p_{\rm ws}}{\nu_{\rm eff}}\left( \eb\eb-\frac{1}{3}\msb{I}\right)\bdbldot\grad \bb{u}_{\rm ws},
\end{equation}
where we have left the turbulent pressure anisotropy ($\Delta p_{\rm tu}$) as-is for brevity. We emphasize that, in the ``ws'' contribution to equation~\eqref{eq:sepdp}, pressure anisotropy is driven by both the wave and the sheared flow (as opposed to being driven by the wave perturbation alone, like in \S\ref{sec:nun0}), because the ``ws'' velocity $\bb{u}_{\rm ws}$ includes $2Ax\ey$ and the magnetic field is not purely azimuthal. 

We can now isolate the evolution of the spiral waves by performing a $z$-average of the dynamical equations. If the magnetic field is isotropically tangled by the turbulence, then the $z$-averaged viscous stress becomes:
\begin{equation}\label{eq:tavg}
    \mnz{\msb{T}_{\Delta p}} = -\mnz{\left( \eb\eb-\frac{1}{3}\msb{I}\right)\Delta p} \approx -\mnz{\left( \eb\eb-\frac{1}{3}\msb{I}\right)\Delta p_{\rm tu}} - \left[\frac{p_{\rm ws}}{5\nu_{\rm eff}} \left( \grad \bb{u}_{\rm ws} +(\grad \bb{u}_{\rm ws})^{\mathsf{T}} - \frac{2}{3}\msb{I}\grad\bcdot\bb{u}_{\rm ws}\right)\right].
\end{equation}
To understand how we arrived at the final term in this expression, first recall that the averaging passes through $\bb{u}_{\rm ws}$ and $p_{\rm ws}$. Then, when equation~\eqref{eq:sepdp} is inserted into equation~\eqref{eq:stress} and averaged, one need only determine $\langle\eb\eb\rangle_z$ and $\langle\eb\eb\eb\eb\rangle_z$ to find the wave/background part of the viscous stress. Borrowing from the linear treatment of magneto-elastic waves in tangled-field configurations and studies of the turbulent dynamo~\citep[e.g.,][]{moffatt85,mk02,hosking20}, we assume that, given many uncorrelated fluctuations over the $z$ dimension, $\langle\eb\eb\rangle_z$ and $\langle\eb\eb\eb\eb\rangle_z$ are isotropic, unity-trace tensors. For tensors of rank 2 and 4, these are $\frac{1}{3}\delta_{ij}$ and $\frac{1}{15}(\delta_{ij}\delta_{nm}+\delta_{in}\delta_{jm}+\delta_{im}\delta_{nj})$ respectively. The resultant viscous stress is then equivalent to the isotropic value suppressed by a factor of 5~\citep{mk02,gruzinov02,gruzinov06}.

With the tangled-field viscous stress estimated, we may address the remainder of the momentum equation. In equation~\eqref{eq:bmom}, the terms containing $\mnz{\rho\delta \bb{u}}$ are simplified further by noting that we are already limited to the large-$\beta$ regime, where density fluctuations from MRI turbulence satisfy $\delta\rho\ll \rho_0$~\citep{salvesen16}. Accordingly, given the assumed linearity of the spiral waves themselves, $\mnz{\rho\delta \bb{u}} \approx \rho_0 \mnz{\delta \bb{u}} = \rho_0\delta \bb{u}_{\rm ws}$. In all, the averaged momentum equation becomes
\begin{equation}\label{eq:smom}
    \rho_0\left( \pD{t}{} + 2Ax\pD{y}{}\right)\delta \bb{u}_{\rm ws} \approx -\grad p_{\rm ws}-2\rho_0\Omega\ez\btimes\delta \bb{u}_{\rm ws}-2A\rho_0 \delta u_{x,\rm ws}\ey + \grad \bcdot \left[\frac{p_{\rm ws}}{5\nu_{\rm eff}} \left( \grad \bb{u}_{\rm ws} +(\grad \bb{u}_{\rm ws})^{\mathsf{T}} - \frac{2}{3}\msb{I}\grad\bcdot \bb{u}_{\rm ws}\right)\right] + \mnz{\mathcal{N}}.
\end{equation}
In \eqref{eq:smom}, we have separated out the ``tu'' viscous stress from equation~\eqref{eq:tavg}, and gathered it with the remaining nonlinear stresses in equation~\eqref{eq:stress} into a term denoted by $\mathcal{N}$ (following H\&P's notation). In \citet{hp09_2}, the authors used ideal MHD simulations of MRI turbulence to demonstrate that the averaged nonlinearity $\mnz{\mathcal{N}}$ is negligible in comparison to all density-wave contributions to the momentum equation. The role of $\mnz{\mathcal{N}}$ may be even weaker in this work: \citet{mk23} showed that Alfv\'enic motions (such as the MRI) interact weakly with compressive dynamics (such as spiral density waves) in high-$\beta$, collisionless plasmas. Motivated by these results, we neglect $\mnz{\mathcal{N}}$ henceforth. 

The only equation remaining to be determined is the evolution equation for the wave-driven isotropic pressure, for which we use \eqref{eq:isop}. It might seem that we can once again apply the same averaging procedure to \eqref{eq:isop} that was used to obtain \eqref{eq:smom}. However, doing so would incorrectly overestimate the tangled-field heat flux. In the presence of a tangled magnetic field, the heat flux is suppressed by the longer path length that particles must travel in order to transport heat from one location to another. According to \citet{malyshkin2001}, this suppression is proportional to the square of the ratio of the magnetic-field correlation length to the temperature-gradient scale, the latter being equivalent to the spiral-wave wavelength, ${\sim} 2\pi H$. Collisionless MRI simulations indicate that the magnetic field is correlated on scales smaller than $H$~\citep[cf. figure 3 of][]{kunz16}, so we expect the heat flux to be suppressed beyond the $\langle\eb\eb\rangle=1/5$ amount by an additional factor $\gtrsim 4\pi^2$. Given such strong suppression, the heat fluxes may be neglected entirely; therefore, the isotropic pressure approximately follows the ideal adiabatic relation $\rmd/\rmd t(p/\rho^{5/3})=0$.\footnote{Even if the parallel heat flux were not suppressed by the additional tangling factor, its primary consequence for spiral-wave evolution would be to smear out temperature fluctuations and enforce an isothermal pressure relation. Including a strong, isotropic heat flux in our calculations does not increase wave damping to the degree that observations would be affected.} 

We now use the adiabatic relation to eliminate $p_{\rm ws}$, and insert the assumption of linear, shearing waveforms \eqref{eq:form} into \eqref{eq:smom} (with $\mnz{\mathcal{N}}$ dropped) and the continuity equation to find a system of ODEs that is readily integrable numerically:
\begin{subequations}\label{eq:blin}
\begin{equation}
    \D{t}{\delta u_{x,\rm ws}}=-\imag\left(k_x(t)+\frac{2A}{5\nu_{\rm eff}}k_y\right)c_{\rm s}^2\frac{\delta \rho_{\rm ws}}{\rho_0}+2\Omega\delta u_{y,\rm ws} - \frac{p_0}{5\rho_0\nu_{\rm eff}} \left(\frac{4}{3}k_x(t)^2\delta u_{x,\rm ws} + k_y^2 \delta u_{x,\rm ws} + \frac{1}{3}k_yk_x(t) \delta u_{y,\rm ws}\right),
\end{equation}
\begin{equation}
    \D{t}{\delta u_{y,\rm ws}}=-\imag\left(k_y+\frac{2A}{5\nu_{\rm eff}}k_x(t)\right)c_{\rm s}^2\frac{\delta \rho_{\rm ws}}{\rho_0}-2(A+\Omega)\delta u_{x,\rm ws} - \frac{p_0}{5\rho_0\nu_{\rm eff}} \left(\frac{4}{3}k_y^2\delta u_{y,\rm ws} + k_x(t)^2 \delta u_{y,\rm ws} + \frac{1}{3}k_yk_x(t) \delta u_{x,\rm ws}\right),
\end{equation}
\begin{equation}
    \D{t}{ \delta \rho_{\rm ws}} = -\imag k_y\rho_0\delta u_{y,\rm ws}-\imag k_x(t)\rho_0\delta u_{x, \rm ws}.
\end{equation}
\end{subequations}
Because of the assumptions of isotropic tangling and weak nonlinearity, we no longer have any direct contribution from the magnetic field. In the absence of rotation and shear then, the waves described by the system~\eqref{eq:blin} become indistinguishable from viscously damped acoustic waves. For this reason, we refer to the spiral density waves in this section as acoustic waves rather than fast waves. We make this distinction solely for the purpose of emphasizing the dominant physics at play, since, in the high-$\beta$ limit, fast waves and hydrodynamic acoustic waves are closely related. We are also only concerned with the ``wave and sheared flow'' fluid quantities now, so the subscript ``ws'' is dropped henceforth.

In the next three sections, we present solutions to the system \eqref{eq:blin} for initially leading spiral waves (\S\ref{sec:awaw}, as a parallel study to \S\ref{sec:nu0} and \S\ref{sec:nun0}). We then discuss the physics involved in the excitation and damping of spiral waves that are converted from vorticity fluctuations (\S\ref{sec:exp}), and finally numerically solve for the full evolution of converted spiral waves as a direct, collisionless analog to the process described by H\&P (\S\ref{sec:vwaw}).

\subsubsection{Initial Acoustic Wave Solutions}\label{sec:awaw}

To isolate the effects of viscous damping due to a background turbulent $\nu_{\rm eff}$ on the propagation of spiral density waves alone, we first perform an analogous study to \S\ref{sec:nu0} and \S\ref{sec:nun0}. As in those sections, we numerically integrate the system \eqref{eq:blin} for the evolution of a leading acoustic wave, which is initialized as a leading perturbation to $u_x$ alone and is evolved from $k_x = -10k_y$ to $k_x = 10k_y$. The scattering rate, motivated in \S\ref{sec:turbb}, is set to $\nu_{\rm eff} = (3/4)\beta\Omega$.

\begin{figure}
\centering
\includegraphics[width=0.74\linewidth]{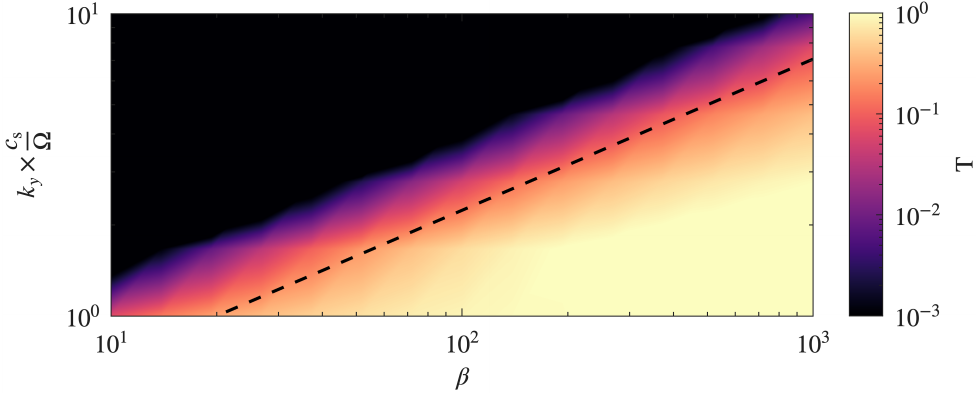}
\caption{Leading to trailing transfer coefficient for spiral waves in a turbulent background magnetic field, assuming $\nu_{\rm eff} = (3/4)\beta\Omega$. Moderate $\beta$ waves are most strongly damped because the viscosity is proportional to $\beta^{-1}$. In the high-$\beta$ limit, waves propagate undamped as they would in an ideal MHD plasma. Note the different axis limits from earlier figures, which ensure that we do not violate the assumptions of Braginskii-MHD. The dashed line represents the WKB prediction for $\rm T=0.1$ obtained via \eqref{eq:freq}.}
\label{fig:twscan}
\end{figure}
Figure \ref{fig:twscan} shows the transfer coefficient $\rm T$ for this acoustic spiral density wave, given the same definition \eqref{eq:transc} employed in \S\ref{sec:azimb}. Here, the range of $\beta$ values begins at 10, which is equivalent to the ratio $|k_x|/k_y$ at the start of the simulation (to ensure the scattering-rate condition for the use of Braginskii-MHD is met).\footnote{The Braginskii requirement $\beta\Omega \gtrsim kv_{\rm th,i}$ is technically violated in the top left corner of figure \ref{fig:twscan}, however we stress that the waves are already strongly damped before one gets to such high $k_yc_{\rm s}/\Omega$ at $\beta=10$. In all other regions of figure \ref{fig:twscan}, $\beta\Omega \gtrsim kv_{\rm th,i}$ is satisfied by the density waves.} For the parameter range shown in figure \ref{fig:twscan}, significant amounts of damping occur for spiral density waves at moderately large values of the plasma $\beta$, even for waves with low azimuthal wavenumber. As $\beta$ increases however, the scattering rate grows and the viscosity vanishes, leading to virtually undamped waves in the very high-$\beta$, low-$k_y$ portion of the plot. Physically, this high-$\beta$ limit is one in which the scattering rate has grown to far exceed the wave frequency, which itself is independent of $\beta$ given the assumptions of \S\ref{sec:turbb}. Accordingly, at such high $\nu_{\rm eff}$, the pressure anisotropy becomes very small, and the viscous damping with it.

To better understand the dependence of the transfer coefficient in figure \ref{fig:twscan} through analytic means, we can employ an identical WKB approach to that used in \S\ref{sec:nu0}. Once again, we assume that the magnitude of the shear $|A|$ is small, in this case $\ll k_y c_{\rm s}$, where $v_{\rm th,i}$ has been replaced by $c_{\rm s}$ as the characteristic wave speed. Given the slow evolution of $\bb{k}$ that follows from this assumption, the waves approximately satisfy the instantaneous dispersion relation at a given $\bb{k}$:
\begin{equation}\label{eq:twdisp}
    \omega^3 + \frac{7\imag}{25}\left(\frac{k^2c_{\rm s}^2}{\nu_{\rm eff}}\right)\omega^2-\left[ k^2c_{\rm s}^2+4\Omega^2+\frac{12}{625}\frac{c_{\rm s}^4}{\nu_{\rm eff}^2}\left(k_x^4+\frac{4}{3}k_y^2k_x^2+k_y^4\right)\right] \omega - \frac{3\imag}{25}\frac{k^4c_{\rm s}^4}{\nu_{\rm eff}} = 0.
\end{equation}
Because \eqref{eq:twdisp} is only third order, the solutions can be obtained exactly, however their full forms are too cumbersome with which to work. Fortunately, the requirement $\nu_{\rm eff}\gtrsim kv_{\rm th,i}\sim kc_{\rm s}$ inherent to Braginskii-MHD guarantees that the damping rate of these acoustic waves will always be small compared to the real part of their frequency, so we once again use the expansion \eqref{eq:asyma} to approximate the real and imaginary parts of the acoustic wave frequency with \eqref{eq:asymp}, finding:
\begin{subequations}\label{eq:freq}
\begin{equation}\label{eq:wr}
    \omega_{r} \approx \pm k_yc_{\rm s}\sqrt{\frac{k^2}{k_y^2}+\frac{4\Omega^2}{k_y^2c_{\rm s}^2}+\frac{12k_y^2c_{\rm s}^2}{625\nu_{\rm eff}^2}\left( \frac{k^4}{k_y^4}+\frac{4k_x^2}{3k_y^2} \right)},
\end{equation}
\begin{equation}\label{eq:wi}
    \omega_i \approx -\frac{7k_y^2c_{\rm s}^2}{50\nu_{\rm eff}}\left(\frac{k^2}{k_y^2}\right) + \frac{3k_y^2c_{\rm s}^2}{50\nu_{\rm eff}}\left(\frac{k^4}{k_y^4}\right)\frac{k_y^2c_{\rm s}^2}{\omega_r^2}.
\end{equation}
\end{subequations}
Given $\Delta p \ll p$ and the corresponding assumption of weak viscosity, the final term in the square root of \eqref{eq:wr} must be small compared to the others. Furthermore, if we assume Keplerian rotation with $\Omega = \frac{4}{3}|A|$, then the assumption $|A| \ll k_y c_{\rm s}$ reduces \eqref{eq:wr} to the simple acoustic wave frequency $\omega_r = \pm kc_{\rm s}$, as expected. When $\omega_r^2=k^2c_{\rm s}^2$ is substituted into \eqref{eq:wi}, we obtain $\omega_i \approx -2k^2c_{\rm s}^2/(25\nu_{\rm eff})$. Given $k_x(t) = -(10+2At)k_y$ and Keplerian shear, we approximate the results of figure~\ref{fig:twscan} by evaluating the integral \eqref{eq:int} with this $\omega_i$ from $t=0$ until $k_x(t) = -10k_y$ to obtain ${\rm T} \approx \exp[ -48k_y^2c_{\rm s}^2/(\beta\Omega^2)]$. We plot the curve that this approximation predicts for $\rm T=0.1$ as a dashed line in figure~\ref{fig:twscan}. The agreement between the numerical and analytic solutions indicates that the lowest-order WKB approximation is effective for capturing the damping of these spiral acoustic waves, and that this agreement holds even when $|A|\sim k_yc_{\rm s}$.

\subsubsection{Expectations: Initial Vortical Wave}\label{sec:exp}

The previous sections have investigated the propagation and damping of collisionless spiral waves initialized as leading fast/acoustic waves. However, the actual process expected to occur in accretion flows involves their swing-induced excitation from fluctuations in the vorticity (\S\ref{sec:hprev}; H\&P, \citet{chagelishvili97}). If we wish to make an effective prediction about observable properties of spiral density waves in non-LTE accretion flows, we must probe the excitation of these waves as well. Before finding numerical solutions for this process, however, let us first consider some of the changes that occur to the governing equations described in H\&P as a result of micro-instability-induced viscosity.

In their work, H\&P found that static fluctuations in the pseudo-potential vorticity (PPV),
\begin{equation}
    \xi \equiv \pD{x}{}(\rho u_y)- \pD{y}{}(\rho u_x) - 2(A+\Omega)\rho,
\end{equation}
which resemble static entropy waves (also known as vortical waves), excite acoustic waves when they undergo a swing. It can be shown that, in the absence of viscosity, the pseudo-potential vorticity is conserved by the system \eqref{eq:blin}. This, in turn, implies that there is no mechanism in the ideal linearized equations to produce the requisite vortical waves (H\&P). When including the turbulent nonlinearities that were dropped to obtain \eqref{eq:blin}, however, \citet{hp09_2} demonstrated that $\xi$ grows slowly compared to the vertical sound crossing time ($H/c_{\rm s}$) due to the underlying MRI turbulence. Therefore, turbulence slowly builds up fluctuations in the PPV, which are otherwise conserved, so that they eventually swing and are converted into spiral density waves.

In the collisionless, turbulent system we describe with \eqref{eq:blin}, the PPV is no longer conserved. Using \eqref{eq:blin} to calculate the evolution equation for fluctuations in the PPV, $\delta \xi \approx \rho_0\partial_xu_y-\rho_0\partial_y u_x-2(A+\Omega)\delta\rho$, we find that
\begin{equation}\label{eq:vorte}
    \D{t}{\delta \xi}= \frac{2Ap_0}{3\rho_0\nu_{\rm eff}}\left(k^2_x(t)-k_y^2\right)\delta \rho - \frac{p_0k^2(t)}{5\rho_0\nu_{\rm eff}}\bigl[\delta \xi+2(A-\Omega)\delta \rho\bigr].
\end{equation}
At $|k_x(t)|\gg |k_y|$, PPV fluctuations, or vortical waves, possess negligible $\delta \rho/\rho_0$, so according to \eqref{eq:vorte} their primary behavior is simply to damp. This is also true when the shear is small. By neglecting the highest-order terms in $|A|/k_yc_{\rm s}\sim \Omega/k_yc_{\rm s}\ll 1$ and inserting our approximation for $\nu_{\rm eff}$, we integrate equation~\eqref{eq:vorte} directly from an initial (negative) $k_{x,0}$ and $\delta \xi_0$ to obtain an estimate of the decay of the vortical wave as it approaches the swing (for Keplerian shear):
\begin{equation}\label{eq:ppvdec}
    \delta \xi(t) \approx \delta \xi_0\exp\left[ -\frac{8}{75}\frac{k_y^2c_{\rm s}^2}{\beta\Omega^2}\left( \frac{k_x(t)-k_{x,0}}{k_y}+\frac{k^3_x(t)-k_{x,0}^3}{3k_y^3} \right) \right].
\end{equation}
Therefore, not only does viscosity damp spiral waves after excitation, it also affects the buildup of vorticity that is necessary for density-wave excitation in the first place. With \eqref{eq:blin} being a linear set of equations, the amplitude of the excited density wave is proportional to the amplitude of $\delta \xi$ when $k_x(t)=0$~(H\&P). So, assuming initially tightly wound vortical modes ($|k_{x,0}| \gtrsim 3k_y\sim 3H^{-1}$, so that the cubic term in \eqref{eq:ppvdec} dominates), we predict that the excitation of spiral density waves is suppressed by a factor $\gtrsim\rme$ whenever 
\begin{equation}\label{eq:predamp}
\beta \lesssim |k_{x,0}/k_y|^3/10.
\end{equation}
Seeing as our Braginskii approximation is valid only when $\beta \gtrsim |k_{x,0}/k_y|$, the smallest value of $\beta$ for which we can reliably predict excitation to be suppressed is $\beta_{\rm min} {\approx} 3$. We emphasize, however, that \eqref{eq:predamp} includes only dissipation incurred prior to the swing, and spiral waves may be suppressed down to $\beta{\sim} 1$ due to damping that occurs after the swing. 

\begin{figure}
\centering
\includegraphics[height=0.21\textheight]{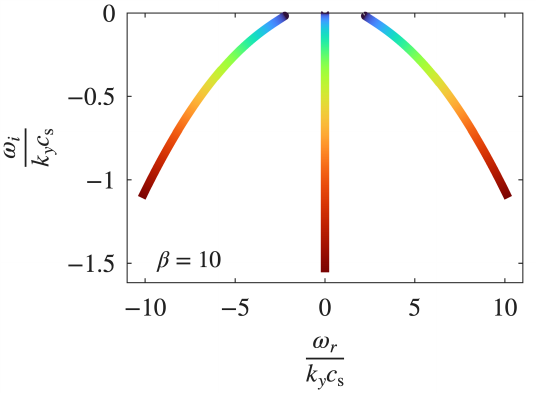}
\includegraphics[height=0.21\textheight]{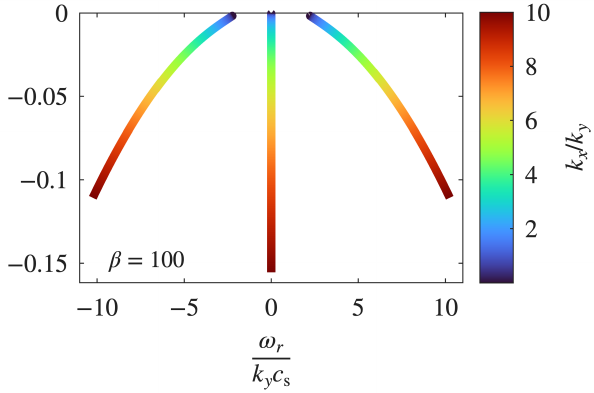}
\caption{Solutions of the dispersion relation of shearing waves in an isotropically tangled magnetic field, for $0 < |k_x| < 10k_y$ at $\beta = 10$ and $100$. In each case, the vortical wave is the branch with no real frequency, while the acoustic spiral wave solutions have a real frequency that is much larger than their damping rate at all wavenumbers. Note the apparent linear dependence of damping rate on $\beta^{-1}$, which explains the variation of $\rm T$ seen in figure \ref{fig:twscan}.}
\label{fig:twdisp}
\end{figure}

There is also the matter of how vortical-acoustic wave conversion (and therefore density wave excitation) occurs in a viscous system. To develop a qualitative understanding of this process, let us probe the linear dispersion relation \eqref{eq:twdisp}, which reveals the relationship between the acoustic- and vortical-wave branches. In figure~\ref{fig:twdisp} we plot all (vortical and acoustic) exact solutions to \eqref{eq:twdisp} for two values of $\beta$ (10 and 100) and $k_yc_{\rm s}=\Omega$.
The color on each branch corresponds to the value of $|k_x/k_y|$ for which the solution was obtained, the horizontal axis corresponds to its real frequency, and the vertical axis to its damping rate. In the spiral wave excitation process, we begin at the bottom of the $\omega_r=0$ vortical wave branch with a leading wave satisfying $|k_x| \gg |k_y|$. As the wave shears, $|k_x|$ decreases and the wave moves upwards along the dispersion branch, which in turn slows the rate of damping. Eventually, when the wave swings ($k_x=0$), the dispersion relations of the vortical wave and the acoustic waves nearly converge. It is at this point where near-resonance is achieved between the acoustic and vortical waves, allowing energy to be exchanged between the coupled branches in a process often referred to as ``linear mode conversion''~\citep[see, e.g.,][]{friedland87,hinkel92,tracy93}. After the swing, when $|k_x|$ begins to increase, the newly generated acoustic wave grows both in frequency and damping rate, yielding a trailing spiral density wave, which, when considering damping that has occurred throughout this conversion process, is lower in amplitude than it would be if viscosity were absent. By comparing the vertical axes limits at $\beta=10$ and $\beta=100$, we see that the damping rate is inversely proportional to $\beta$, or alternatively, linearly proportional to magnetic energy density. This is no surprise given that the effective viscosity in our micro-unstable plasma is $\mu_{\rm eff} \equiv p/\nu_{\rm eff} \propto B^2/\Omega$.

\subsubsection{Initial Vortical Wave Solutions}\label{sec:vwaw}

\begin{figure}
\centering
\mbox{\hspace{6em}$(a)$\hspace{0.8\textwidth}}
\includegraphics[width=0.84\linewidth]{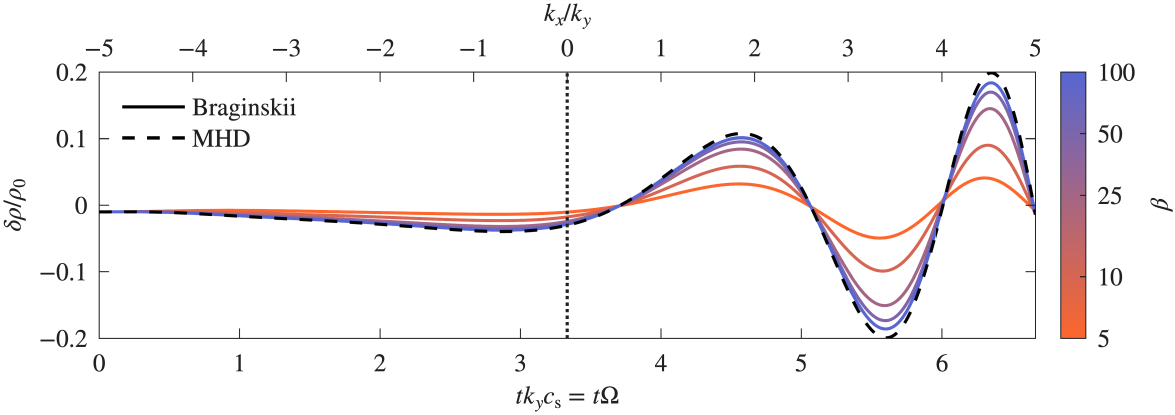}\\ 
\mbox{\hspace{6em}$(b)$\hspace{0.8\textwidth}}
\includegraphics[width=0.84\linewidth]{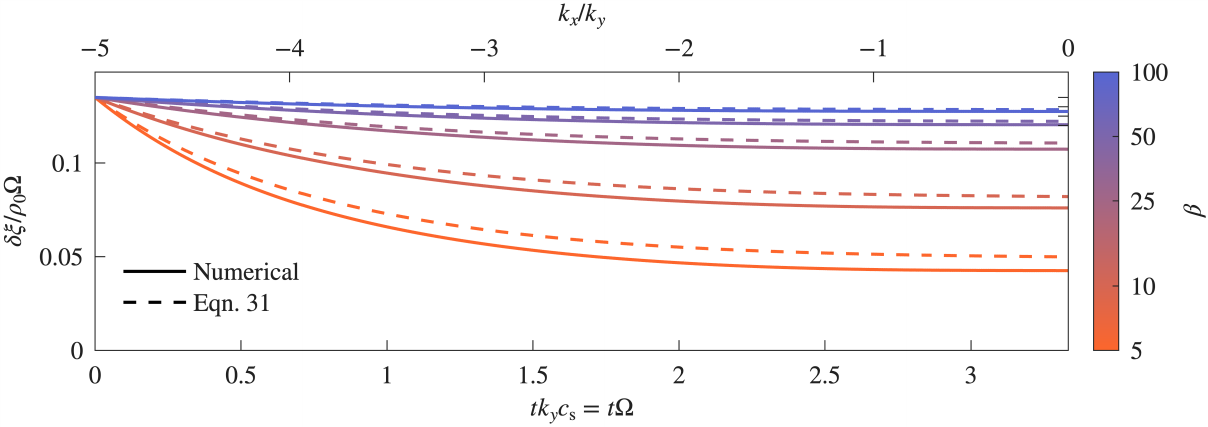}
\caption{(a) The evolution of leading vortical-wave density fluctuations with $k_y = H^{-1}$ as they are converted into spiral density waves when they swing (at a time denoted by the vertical dotted line). The dashed line is the MHD limit of this process, as studied in H\&P. The corresponding $\beta$ for each solution is indicated by the respective tick mark on the color scale, the most damped being $\beta{=}5$, and the least damped being $\beta{=}100$. (b) Pre-swing evolution of the PPV fluctuations for the Braginskii solutions shown in panel (a) (solid lines), compared to the approximation in the limit of $k_y \gg |A|/c_{\rm s}$ \eqref{eq:ppvdec} (dashed). Significant damping of the PPV occurs prior to excitation, which is well-described by the analytic prediction \eqref{eq:ppvdec} even though these waves possess $k_y\sim |A|/c_{\rm s}$.}
\label{fig:ivort}
\end{figure}

With our expectations for the spiral wave excitation process laid out, we now solve the system \eqref{eq:blin} numerically given appropriate initial conditions. We seed an initially leading vortical wave at $k_{x,0} = -5k_y$ with $\delta u_y=0.1\imag v_{\rm th,i}$ and the eigenvector relationships for the inviscid vortical wave: $\delta u_x = -k_y\delta u_y/k_x$ and $\delta \rho = -2\imag\rho_0 \,\Omega \, \delta u_y/(k_xc_{\rm s}^2)$. The choice of smaller initial $|k_x|$ is made here to enable us to explore values of $\beta$ down to 5, while not violating the assumptions of Braginskii-MHD. In figure~\ref{fig:ivort}(a), the numerical solutions for $\delta \rho(t)$ are provided for $\beta \in [5,100]$, with $k_y c_{\rm s} = \Omega$ and $\nu_{\rm eff} = (3/4)\beta \Omega$. The tick marks in the color scale correspond to the exact values of $\beta$ for the solutions given.
The dashed line, which is the inviscid (ideal MHD) solution, shows that the initially leading vortical wave possesses a small density fluctuation that grows as the swing time is approached (vertical dotted line). After the swing (and the vortical-acoustic conversion), the density fluctuation oscillates, growing in amplitude from action conservation. When the effects of viscosity are included from finite $\nu_{\rm eff}$ (solid lines), one can see that damping leads to a range of final density-wave amplitudes for varying $\beta$, with the lowest values of $\beta$ featuring waves that are most heavily suppressed. Amplitude growth by action conservation immediately after the swing is especially weak for the two lowest-$\beta$ Braginskii solutions, which correspond to $\beta \leq 10$. These strongly damped density waves are also excited at low amplitudes, implying that the PPV fluctuations necessary for density wave excitation are being damped prior to the swing time. In figure \ref{fig:ivort}(b), we show the evolution of the leading vortical-wave PPV perturbations ($\delta \xi$) for the same Braginskii solutions presented in panel (a) (solid), alongside the prediction for the evolution of $\delta \xi$ given by \eqref{eq:ppvdec} (dashed). Indeed, we find that the lowest-$\beta$ vorticity fluctuations are damped by more than a factor of two from their initial value before they reach the swing time. We also find that the approximation \eqref{eq:ppvdec}, which was derived in the limit of $k_y \gg |A|/c_{\rm s}$, holds well even though these waves possess $k_y \sim |A|/c_{\rm s}$.

\begin{figure}
\centering
\includegraphics[width=0.76\linewidth]{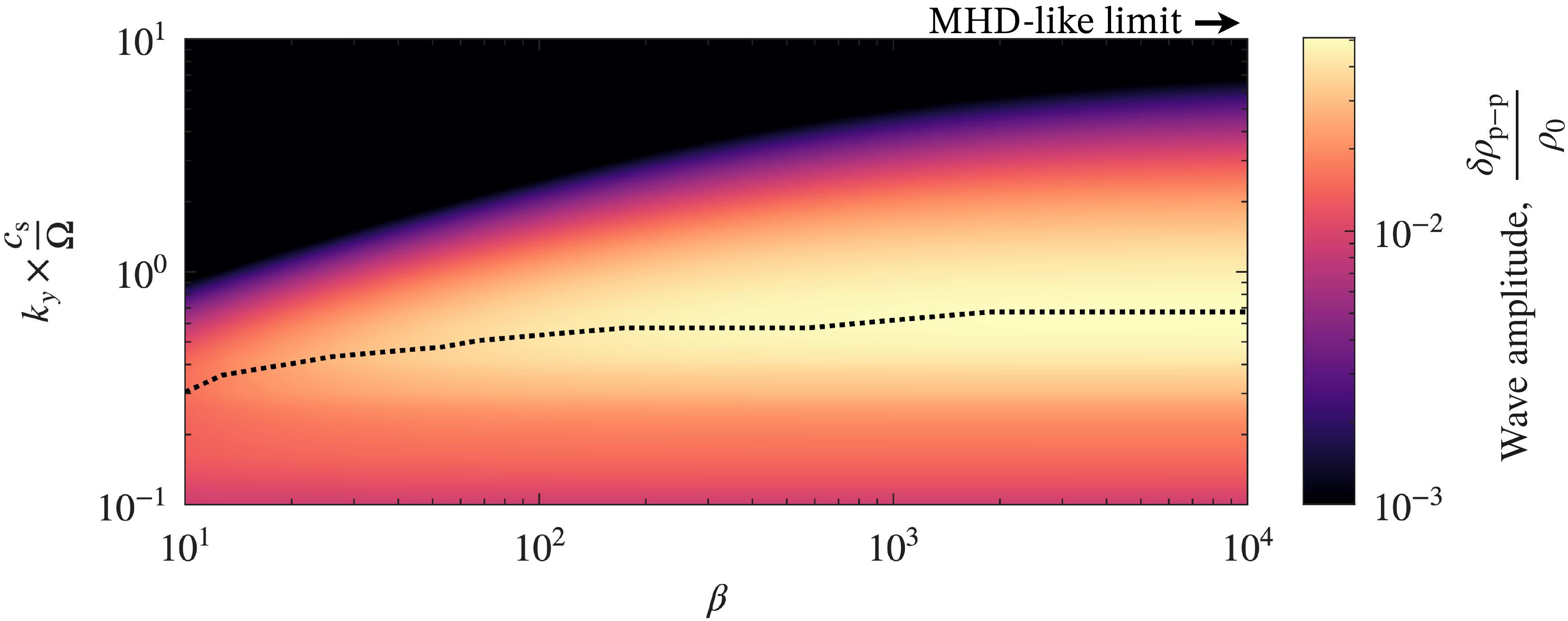}
\caption{Peak-to-peak amplitude of spiral density waves generated by an initial vortical perturbation $\delta \xi = 0.01 \rho_0\Omega$, as a function of $k_y$ and $\beta$, assuming $k_{x,0} = -10k_y$. The dotted line tracks the peak $k_y$ for excitation as a function of $\beta$. Viscous damping leads to a skew of the peak excitation wavenumber toward smaller $k_y$ at moderate $\beta$. At very high $\beta$, the peak $k_y$ settles near the ${\sim} H^{-1}$ prediction of H\&P because the viscosity is microphysically regulated to very small values. As a result, non-LTE spiral waves at moderate $\beta$, such as those likely to arise in low-luminosity accretion flows, feature lower amplitudes and azimuthal wavenumbers than their ideal-MHD counterparts (approximated here by the asymptotically high-$\beta$ limit).}
\label{fig:excite}
\end{figure}
In figure \ref{fig:ivort}, we only studied waves with $k_y = H^{-1}$, because wavenumbers near this value were predicted by H\&P to be excited at the largest amplitudes. However, it is worth probing whether this remains the case in the presence of viscous damping. H\&P demonstrated that, although the excited wave amplitudes peak near $k_y \sim H^{-1}$, there is a finite width to the distribution of excited wavenumbers that extends symmetrically to higher and lower $k_y$. In the collisionless system, viscous damping preferentially damps large-$k_y$ waves (because the ratio of their shear time to wave period, and therefore damping time, is longer). This preference should induce an asymmetry in excited wave amplitudes that skews the distribution to lower $k_y$. Given the dependence of the damping strength on $\beta$, this skew should become more prominent with decreasing $\beta$.

In Figure~\ref{fig:excite}, we test these expectations by initializing leading vortical waves with $k_{x,0} = -10k_y$, and evolving them through the conversion process until the wavenumber is reversed ($k_{x,0} = 10k_y$). Rather than initializing a fixed $\delta u_y$, here we initialize a fixed PPV fluctuation, $\delta \xi = 0.01 \rho_0\Omega$. This choice is motivated by \citet{hp09_2}, where the authors found that at low $k_yc_{\rm s}/\Omega$ the spectrum of PPV supplied by MRI turbulence is approximately flat. The shading in figure~\ref{fig:excite} shows the final peak-to-peak amplitude of the resulting spiral density wave given the (fixed) initial PPV perturbation. At moderate $\beta$ ($\lesssim 30$), where damping is stronger, the distribution of excited wave amplitudes is not only decreased all around, but skewed toward lower $k_y$ as expected. The dotted line tracks the $k_y$ of the most strongly excited wave at each $\beta$, finding that at $\beta=10$ the peak azimuthal wavenumber is lower than the MHD equivalent (which is effectively the same as $\beta \gtrsim 10^3$ in figure~\ref{fig:excite}) by a factor $\gtrsim 2$. Importantly, however, this result is dependent on the initial ratio $|k_{x,0}/k_y|$: the peak $k_y$ for excitation can be decreased further by initializing a leading vortical wave with a larger value of $|k_{x,0}/k_y|$, so that the wave has more time to damp before and after the swing. Alternatively, the low-$k_y$ skew can be weakened by initializing a wave with $|k_{x,0}|\lesssim |k_y|$. In H\&P, the authors seeded $\delta \xi$ at values of $|k_{x,0}/k_y|$ that were so large (${\sim} 100$) that the vortical waves would damp entirely prior to the swing time in the viscous framework studied here. Whether such radial leading waves are the most accurate choice for the turbulent source of $\xi$ is presently unclear. Therefore, for \textit{precise} predictions of the change in the dominant $k_y$ at which spiral density waves are excited, $|k_{x,0}/k_y|$ for the PPV fluctuations will have to be constrained using future collisionless accretion-flow turbulence simulations.

\section{Discussion: Observational and Modeling Implications}\label{sec:disc}

Spiral density waves are predicted to play a significant role in determining the morphology and amplitude of fluctuations in the emission region of a black-hole accretion flow. We have demonstrated that the excitation and propagation of spiral density waves in collisionless plasmas can be suppressed. This finding has consequences for accretion-flow observables across a broad range of low-luminosity systems, including those relevant to the EHT project. There are, of course, many plasma-independent effects that determine the actual appearance of black-hole images and mock observations from GRMHD simulations~\citep{eht22models}. We therefore do not make any precise statements here about what the exact light curve variabilities or visibility amplitudes should be from accretion-flow observations. We can, however, make predictions about how such observations ought to differ from those predicted by GRMHD simulations without Landau/Barnes damping or microphysically regulated viscous stresses. 

We first consider the amplitude of the spiral density waves, which we associated with non-axisymmetric fluctuations in emitted synchrotron power near the black-hole event horizon. We find that spiral density wave amplitudes are significantly suppressed by viscous stresses at moderately large values of the plasma $\beta$ ($\in [3,30]$).
Given the potential role of density waves in determining the variability of accretion-flow emission, such damping may have important observational consequences. In particular, the absence of non-LTE damping mechanisms in GRMHD models may help explain why the EHT library, on average, overestimated the variability of Sgr A$^\ast$.\footnote{Certain models of the Sgr~A$^\ast$ accretion flow, referred to as ``magnetically arrested''~\citep[MAD;][]{Igumenshchev2003,narayan03,Tchekhovskoy2011}, are often characterized as being low $\beta$ ($\beta < 1$). One might therefore conclude that these models fall outside of the range of $\beta$ in which we predict significant suppression of spiral waves ($\beta \in [3,30]$). However, in such models the low-$\beta$ state is achieved close to the Sgr~A$^\ast$ event horizon, and beyond ${\sim} 10 \, r_g$ the plasma $\beta$ can grow to be $>1$~\citep{dhruv25}. Because spiral waves are thought to originate outside the emission region~\citep{conroy25}, which itself is constrained to be within ${\sim}5\, r_g$~\citep{eht22_1}, spiral-wave suppression may be significant even in MAD models of the Sgr~A$^\ast$ accretion flow.} Thus far, only one global investigation of weakly collisional accretion has been performed with the goal of modeling Sgr~A$^\ast$~\citep{dhruv25}, and their results indicate that variability is indeed reduced in the presence of viscous dissipation. In their case, the observed decrease in variability was modest, however this may be a consequence of the fact that they assumed there was an additional (non-micro-instability driven) source of scattering present everywhere \textit{in addition} to that which results from micro-instabilities. Any additional scattering reduces the impact of viscous-damping effects on spiral density waves (given that the viscosity goes as $\mu_{\rm eff} \propto \nu_{\rm eff}^{-1}$), and therefore limits the reduction in variability.
We therefore predict that a truly collisionless approach with a self-consistent treatment of micro-instability scattering will feature weaker spiral density waves and a further reduction in variability. This is, in fact, already supported by earlier simulation results studying MRI turbulence using a hybrid-kinetic shearing-box approach~\citep{kunz16}, which naturally captures the exact scattering rate of micro-instabilities because the instabilities and the particles they scatter are simulated directly. In their simulation, which possessed $\langle\beta^{-1}\rangle \sim 0.15$, large-scale trailing density waves were nearly absent, only being detectable after vertical averaging, and even then having amplitudes $\delta\rho/\rho \lesssim 0.05$ (see their fig.~4). By comparison, the density waves seen in the MHD simulations of \citet{hp09_2} had amplitudes ${\approx}4$ times larger. The suppression that \citet{kunz16} measured in spiral-density-wave amplitudes may also have been reinforced by the fact that, at such low $\beta$, the peak $y$-wavelength for density wave excitation could have been stretched beyond the box length that they employed in the azimuthal direction ($L_y = 4H$).

Another implication of this work concerns EHT observations that are yet to come: emission-region pattern speeds~\citep{conroy23}. Improvements to the EHT will soon make it possible to construct movies of the accretion flows around Sgr~A$^\ast$ and M87$^\ast$~\citep{doeleman19,johnson19}. Given the asymmetric emission already present in still frames of these sources, a time sequence of such images would reveal the movement of any non-axisymmetric features (more precisely their rotation). In movies generated from GRMHD simulations, it has been found that these patterns precess at a rate well below the Keplerian speed at their radius, in other words, the observed evolution is not the result of rigid rotation of a particular region of the flow. Instead, the frequency of this rotation has been shown to coincide with the pattern speed of spiral fast/acoustic waves, $\Omega_{\rm p} \equiv \Omega_0-\omega/m$, where $m=k_yr_0$ is the azimuthal mode number, $\Omega_0$ is the orbital frequency at $r_0$, and the negative sign on $\omega$ reflects the wave's inward propagation~\citep{conroy23}. It is therefore natural to question whether collisionless physics leads to new expectations for the $\Omega_{\rm p}$ of these spiral waves via changes to $m$, $\omega$, and $\Omega_0$. Let us first consider changes in the azimuthal mode number $m$. Given that $m$ is proportional to $k_y$, figure \ref{fig:excite} implies that lower values of $m$ are preferred by collisionless spiral density waves than are preferred by MHD ones. Assuming, for the moment, that the detected waves are excited at the same radius when viscosity is included as they are in its absence, we expect that GRMHD simulations overestimate the pattern speed of emission-region structures, and that they do so more significantly if the plasma $\beta$ is moderate or low. This statement does, however, rely heavily on the assumption that the region from which these spiral waves originate is unchanged by collisionless physics. Any such change will affect the orbital frequency at excitation $\Omega_0$, which would directly modify the pattern speed. In MHD, the radius at which these waves are excited is set by a balance of several factors, including (but not limited to) where the production of vorticity is most efficient and how the wave amplitude changes with radius. We cannot address how these details might change due to collisionless effects in our limited analysis, so fully nonlinear simulations are necessary to resolve this area of uncertainty in our prediction of lower $\Omega_{\rm p}$. As for the wave frequency $\omega$, it too is determined by the radius of excitation. When moving \textit{with} a wave packet inward toward the emission region, geometrical optics dictates that the wave frequency $\omega$ remains unchanged. Hence, the local properties of the flow at excitation, which depend on $r_0$, determine $\omega$. Lastly, it is also possible that, given strong viscous damping, spiral density waves become so low amplitude that they no longer dominate the morphology of the emission region. In this case, the above conclusion that pattern speeds should be overestimated by GRMHD would change entirely, instead simply becoming uncorrelated with spiral density wave excitation. Unfortunately, to address the consequences that possibility holds for the time evolution of black-hole emission, one needs to quantify all of the other potential sources of emission region variability, which is beyond the scope of this work.

\section{Conclusion}\label{sec:conc}

We have presented an investigation of the excitation, propagation, and damping of spiral density waves in collisionless accretion-flow plasmas, in which we consider the roles played by a variety of non-LTE effects and physical setups. In a background magnetic field that is laminar and azimuthal, spiral density waves satisfying $k_yv_{\rm th,i} \gtrsim 10\Omega$ were found to be strongly suppressed by Landau/Barnes damping. Waves at lower $k_y$, however, such as those most strongly excited by MRI turbulence ($k_yv_{\rm th,i}\sim \Omega$), were nearly unaffected by Landau/Barnes damping. When scattering from wave-generated micro-instabilities was included, the resultant viscous damping increased spiral-wave dissipation, even at $k_yv_{\rm th,i}\sim \Omega$. The strongest dissipation of spiral density waves was found to occur in a turbulent background magnetic-field configuration, where micro-instability scattering is primarily driven by the turbulence rather than the waves themselves. At moderately large values of $\beta$ ($\in[3,30]$), we found that spiral density waves are not only viscously damped after excitation, but also that the turbulent fluctuations necessary to excite them in the first place can be suppressed. Given the importance of micro-instability scattering in damping spiral waves, we conclude that suppression is broadly expected to occur in high-$\beta$ accretion flow plasmas, but not low-$\beta$ ones. Furthermore, because spiral density waves perturb the the rate of synchrotron emission from the accreting plasma, these results are expected to impact observations of optically thin, low-luminosity accretion flows. Specifically, micro-instability-mediated viscous damping is predicted to reduce both emission variability and the pattern speed of emission-region features relative to corresponding LTE models. These results therefore demonstrate that micro-physical plasma processes can indirectly impact the observable structure of hot accretion flows, and advocate for future modeling efforts that employ well-motivated closures for non-LTE physics, so that tighter constraints may be placed on the parameter regimes of the accretion flows themselves.

Being a study of spiral density waves in the relatively isolated shearing sheet, there are obvious limitations to this work. Clearly, both general and special relativistic effects play an enormous role in determining the actual morphology of the emission region, at least given some underlying structure inherent to the plasma itself. It is most prudent, however, to consider those effects that we have neglected even in our local approximation to the entire spiral wave generation/propagation process. The most significant of these limitations is the neglect of nonlinear effects (beyond self-generated micro-instabilities). In reality, because spiral density waves are fast/acoustic waves, they are expected to steepen into shocks within roughly $t_{\rm shock} \sim (kc_{\rm s}\delta \rho/\rho_0)^{-1}$ of being excited~\citep{hp12,mks23}. This sets an effective upper limit on the amount of time during which such waves linearly damp. An order-of-magnitude estimate of this time, given $\delta \rho/\rho_0 = 0.1$ and $k_x \sim k_y$, is $tk_yc_{\rm s} \sim 10$ after the swing time. From figure \ref{fig:ivort}, it is clear that this is still enough time to see significant changes in the spiral wave amplitude from damping. Actual differences, however, may be even larger, because viscosity can slow steepening, and, if strong enough, may halt steepening altogether. Carefully designed nonlinear simulations will better constrain the additional differences that occur as a result of this discrepancy in steepening rates. 

After the assumption of linearity, our treatment of the scattering rate is the next-most immediate area in which improvement could be made. In an analytic model, it is difficult to improve accuracy beyond the approach used here unless a multi-scale method can be developed that accurately describes the saturated state of the intermediate-scale MRI turbulence, and therefore its rate of strain. Given that MRI turbulence is an exceptionally active area of research and the general applicability of many such results is still under careful scrutiny, a simulation-based approach is more favorable. The two most physically realistic numerical approaches for such a problem would either be fully- or hybrid-kinetic~\citep{riquelme12,hoshino2015,kunz16,inchingolo2018,sandoval2024}, or a fluid-based approach employing the Chew-Goldberger-Low (CGL) equations~\citep{sharma06,squire17_mri}. The former options, although based on first principles, are computationally expensive, especially if substantial scale separation is desired not only between the spiral wave and the MRI turbulence, but also between the MRI turbulence and the ion-Larmor scale. The CGL fluid approach is much more affordable, but compromises on the accuracy of the heat fluxes, which above all, are responsible for Landau/Barnes damping. If spiral wave behavior is the primary goal, this may be an acceptable compromise, given that accurate viscous stresses are of greater concern to the damping of spiral density waves than Landau/Barnes damping. Nonetheless, the CGL model by itself does not accurately model the excitation and saturation of micro-instabilities, nor their feedback on $\Delta p$, so a well-crafted closure must be employed that realizes the expected rate of scattering from \eqref{eq:scat}. 

Finally, in our derivation of the governing equations for shearing waves in a turbulent magnetic field, we made the simplifying assumption that the field is tangled isotropically. Realistic MRI turbulence, however, does not produce isotropically tangled magnetic fields. If it did, the stresses that drive angular momentum transport---a crucial ingredient of accretion---would be negligible~\citep{balbus94,hawley95,brandenburg96}. Therefore, the equations provided here represent a limiting case of waves in extremely tangled fields. To relax this assumption of isotropy, modifications can be made to the approximations used for $\langle \eb\eb\rangle$ and $\langle \eb\eb\eb\eb\rangle$. For $\langle \eb\eb\rangle$, the $xy$ component of an updated prescription could be estimated from the Maxwell stress $\langle B_xB_y\rangle$ that is often reported in simulations (see, for example, the references to collisionless MRI turbulence cited throughout this work). Second-order correlations among other components of the magnetic field, however, are not often reported. Similarly, the fourth-order correlation, $\langle \eb\eb\eb\eb\rangle$, is not closely related to any quantity that has yet been constrained in MRI turbulence simulations. These additional correlations can, however, be measured in future simulations of the collisionless MRI, enabling a straightforward $\mathcal{O}(1)$ update to the present work.

\section*{Acknowledgements}

The authors thank M.~Begelman, T.~E.~Foster, C.~F.~Gammie, H.~Hakobyan, C.~Hamilton, and J.~M.~Stone for helpful discussions that improved this work. Support for S.~M. was provided by NASA Astrophysics Program grant 80NSSC22K0828. M.~W.~K. was supported in part by NSF CAREER Award No.~1944972. J.~D. was supported in part by NSF award AST-2307983.

\section*{Data Availability}

The data underlying this article will be shared on reasonable request to the corresponding author.




\appendix

\section{Derivation of the shearing sheet Kinetic MHD Equations}\label{app:kmhd}

Here we present a derivation of the shearing-sheet terms that are present in \eqref{eq:kkin} and \eqref{eq:kmom}. For brevity, we only describe in detail the manipulations necessary to obtain the shearing-sheet contributions to kinetic MHD, leaving the derivation of ``standard'' kinetic MHD terms out. The full derivation of kinetic MHD is involved, and has been presented in many prior works, such as \citet{kulsrud64}, \citet{kulsrud83}, and \citet{kulsrudbook}; we encourage interested readers to consult these references.

Our starting point is the collisionless kinetic (Vlasov) equation for a particle species $s$ that represents either electrons or ions:
\begin{equation}\label{eq:vlas}
    \pD{t}{f_s}+ \grad \bcdot (\bb{v} f_s) + \pD{\bb v}{} \bcdot \left( \bb{a}_s(t,\bb{r},\bb{v})f_s\right) = 0.
\end{equation}
Here, $\bb{v}$ represents the particles' velocities and $\bb{a}_s$ their accelerations. Focusing exclusively on the forces unique to the local shearing-sheet approximation, the latter are provided by Hill's equations~\citep{Hill1878,glb65}:
\begin{subequations}\label{eq:hill}
    \begin{equation}
        a_{x,s} = -4\Omega_0 A_0x+2\Omega_0 v_y, 
    \end{equation}
    \begin{equation}
        a_{y,s} = -2\Omega_0 v_x.
    \end{equation}
\end{subequations}
As in the main text, $x$ represents the local radial coordinates and $y$ represents the local azimuthal coordinate; $\bb{\Omega}(r_0) = \Omega_0\ez$ is the rotation frequency and $A_0$ is Oort's first constant. It is therefore implied that the $\bb{v}$ are the particles' velocities in the rotating frame described by \eqref{eq:hill}. With the addition of magnetic ($\bb{B}$) and electric ($\bb{E}$) fields, the total acceleration appearing in \eqref{eq:vlas} is then given by
\begin{equation}
    \bb{a}_s = \frac{q_s}{m_s}\left( \bb{E} +\frac{\bb{v}\btimes\bb{B}}{c}\right) -\grad\psi_{\rm eff} - 2\Omega_0\ez \btimes \bb{v},
\end{equation}
where we have introduced the effective gravitational potential $\psi_{\rm eff} = 2\Omega_0A_0x^2$, the particle charges ($q_s$) and masses ($m_s$), and the speed of light ($c$).

We now transform the Vlasov system to a frame that is co-moving with \textit{all} bulk velocities (not just rotation), so that the resultant $f_s$ possesses no first-order velocity moment. This is accomplished by transforming from the total rotating-frame particle velocity $\bb{v}$ to the peculiar velocities $\bb{w}_s \equiv \bb{v}-\bb{u}_s$, where $\bb{u}_s = n_s^{-1}\int \rmd^3v\,\bb{v} f_s$ is the rotating-frame bulk-flow velocity and $n_s = \int \rmd^3v\, f_s$ is the number density. In this transformation, the time, space, and velocity gradients are modified according to 
\begin{equation}
    \pD{t}{} \rightarrow \pD{t}{} - \pD{t}{\bb{u}_s}\bcdot \pD{\bb{w}_s}{}, \quad \grad \rightarrow \grad - (\grad\bb{u}_s)\bcdot\pD{\bb{w}_s}{}, \quad {\rm and} \quad \pD{\bb{v}}{} \rightarrow \pD{\bb{w}_s}{}.
\end{equation}
Substituting these derivatives and eliminating $\bb{v}$ in favor of $\bb{w}_s$, we obtain
\begin{equation}\label{eq:nou}
    \D{t_s}{f_s}+ \bb{w}_s\bcdot\grad f_s + \left( \frac{q_s}{m_s}\frac{\bb{w}_s\btimes \bb{B}}{c} - 2\Omega_0\ez\btimes \bb{w}_s + \bb{a}_{s,\rm f} - \bb{w}_s \bcdot \grad \bb{u}_s \right)\bcdot \pD{\bb{w}_s}{f_s} = 0,
\end{equation}
where $\rmd/\rmd t_s = \partial/\partial_t +\bb{u}_s\bcdot \grad$ is the convective derivative for species $s$ and
\begin{equation}
    \bb{a}_{s,\rm f} = \frac{q_s}{m_s}\left( \bb{E} + \frac{\bb{u}_s\btimes \bb{B}}{c}\right) - \grad \psi_{\rm eff} - 2\Omega_0\ez\btimes \bb{u}_s - \D{t_s}{\bb{u}_s}
\end{equation}
is a collection of terms referred to as the ``fluid'' acceleration, for the fact that they do not depend on $\bb{w}_s$. Now, rather than the kinetic equation being one that is solved directly for $f_s$ from which $\bb{u}_s$ can be calculated, we have a system in which $\bb{u}_s$ is a field that must be determined separately so that $f_s$ can be evolved. It is therefore necessary to obtain an evolution equation for $\bb{u}_s$.

Taking the $m_s\bb{w}_s$ moment of \eqref{eq:nou} (and noting that $\int \rmd^3w_s\,\bb{w}_s f_s=0$ by construction), the momentum equation is obtained:
\begin{equation}
    m_s n_s\D{t_s}{\bb{u}_s} = -\grad\bcdot \msb{P}_s + q_sn_s\left(\bb{E}+\frac{\bb{u}_s\btimes \bb{B}}{c}\right) - m_sn_s\grad \psi_{\rm eff} - 2\Omega_0 m_sn_s\ez\btimes \bb{u}_s.
\end{equation}
Here $\msb{P}_s = m_s\int\rmd^3 w_s \,\bb{w}_s\bb{w}_sf_s $ is the pressure tensor for each species. From here, the standard MHD assumptions are applied, save for the assumption of LTE: quasi-neutrality ($\sum q_sn_s=0$), negligible electron inertia ($m_{\rm i} \gg m_{\rm e}$), long length scales ($l \gg r_{\rm L}$), and long time scales ($\omega \ll \omega_{\rm L}$). The outcome for the ion momentum equation is typical of MHD except for the presence of the shearing-sheet terms and the fact that the ion pressure remains anisotropic (as in \S\ref{sec:eqs}, the electrons are taken to be cold and therefore make no contribution to the total thermal pressure):
\begin{equation}\label{eq:imom}
    m_i n_i\D{t_i}{\bb{u}_i} = -\grad\bcdot \msb{P}_i -\frac{\grad B^2}{8\pi} + \frac{\bb{B}\bcdot\grad\bb{B}}{4\pi} - m_in_i\grad \psi_{\rm eff} - 2\Omega_0 m_in_i\ez\btimes \bb{u}_i.
\end{equation}
The resultant induction equation for $\bb{B}$ is also standard: 
\begin{equation}\label{eq:indv}
    \D{t}{\bb{B}} = \bb{B}\grad\bcdot\bb{u}_{\rm i} - \bb{B}\bcdot\grad\bb{u}_{\rm i}. 
\end{equation}
There now exists a closed set of equations (\eqref{eq:indv}, \eqref{eq:imom}, and \eqref{eq:nou} with $s=\rm i$) that do not require any information about the (cold) electron distribution function. Henceforth, we drop all ``i'' subscript from the ion quantities.

 Let us now return to the kinetic equation, where our remaining goal is to perform a gyro-average so that we may eliminate any dependence on the gyro-phase angle ($\phi$) from our equations. For convenience, we first introduce an effective electric field,
\begin{equation}
 \bb{E}_{\rm eff} = \bb{E} - \frac{m}{e}\left( \grad \psi_{\rm eff} + 2\Omega_0\ez\btimes \bb{u} \right),
\end{equation}
which absorbs the ``fluid'' shearing-sheet terms into the standard electric field. Now, all of the remaining manipulations that apply to $\bb{E}$ can be applied to $\bb{E}_{\rm eff}$ one-to-one. Consequently, the only new term that we need to gyro-average here is the peculiar Coriolis flux:
\begin{equation}
    F_{\rm C,p} \equiv - (2\Omega_0\ez\btimes \bb{w}) \bcdot \pD{\bb{w}}{f}.
\end{equation}
To make the gyro-averaging process easier, we first exchange the full peculiar velocity $\bb{w}$ in favor of its magnitude ($w$), its projection along $\eb$ ($w_\|$), and the gyro-phase angle ($\phi$). This transforms the $\bb{w}$ derivative of $f$ into
\begin{equation}
    \pD{\bb{w}}{f} \rightarrow \frac{\bb{w}}{w}\pD{w}{f} + \eb\pD{w_\|}{f}.
\end{equation}
Applying this to $F_{\rm C,p}$, we find that only the parallel gradient survives:
\begin{equation}
    F_{\rm C,p} = -2\Omega_0 (\ez \btimes \bb{w}) \bcdot \left( \frac{\bb{w}}{w}\pD{w}{f} + \eb\pD{w_\|}{f}\right) = -2\Omega_0 (\ez \btimes \bb{w})\bcdot \eb\pD{w_\|}{f}.
\end{equation}
We now average this expression over the gyro-angle ($\langle\rangle_\phi$), making use of the fact that the fast-gyrofrequency assumption ($\omega \ll \omega_{\rm L,i}$) makes $f$ gyrotropic, and that $\mng{\bb{w}} = w_\|\eb$. The result is simply that $\langle F_{\rm C,p}\rangle_\phi=0$ because $\mng{(\ez\btimes\bb{w})\bcdot \eb} = (\ez\btimes w_\|\eb)\bcdot \eb = 0$. 

With $\langle F_{\rm C,p}\rangle_\phi=0$, the remainder of the derivation is trivial. We take the standard drift-kinetic equation from kinetic MHD, and substitute $\bb{E}_{\rm eff}$ for the electric field to obtain
\begin{equation}
     \bigD{t}{f} + \bigD{t}{\ln B}\frac{w_\perp}{2}\pD{w_\perp}{f} - \biggl[ \eb\bcdot \biggl(\bigD{t}{\bb{u}} - \frac{e}{m}\bb{E} + 2\Omega_0\ez\btimes\bb{u} + 4A_0\Omega_0 x\ex\biggr)+\frac{w_\perp^2}{2}\nabla_\|\ln B \biggr]\pD{w_\|}{f} = 0.
\end{equation}
One final technical detail concerns the ion pressure tensor $\msb{P}_{\rm i}$. Because $\mng{\bb{ww}} = (w_\perp^2/2)(\msb{I}-\eb\eb) +w_\|^2\eb\eb$, the pressure tensor for a gyrotropic distribution is simply $\msb{P} = p_\perp(\msb{I}-\eb\eb) +p_\|\eb\eb$. With this simplification, the momentum equation \eqref{eq:imom} becomes exactly that which we use in the main text.


\bsp	
\label{lastpage}
\end{document}